\documentclass{jfm}

\usepackage{graphicx}
\usepackage{comment} 
\usepackage{newtxtext}
\usepackage{newtxmath}
\usepackage{natbib}
\usepackage{hyperref}
\hypersetup{
	colorlinks = true,
	urlcolor   = blue,
	citecolor  = black,
}

\newcommand{\RomanNumeralCaps}[1]
\linenumbers

\usepackage{pgfplots}
\pgfplotsset{compat=1.15}
\usepackage{subcaption}

\usepackage{tikz}
\usetikzlibrary{calc}

\usepackage{booktabs}

\newcommand{\walberla}{\textsc{waLBerla}}
\newcommand{\mono}{\texttt{mono}}
\newcommand{\polya}{\texttt{poly}-10}
\newcommand{\polyb}{\texttt{poly}-50}
\newcommand{\polyc}{\texttt{poly}-100}
\usepackage{xcolor}

\graphicspath{{./figures/}}

\usepackage[colorinlistoftodos, textwidth=2cm]{todonotes}
\newcommand{\todoCR}[1][]{\todo[color=yellow, size=\footnotesize, author={Christoph}, inline]}

\newcommand{\todoBV}[1][]{\todo[color=cyan, size=\footnotesize, author={Bernhard}, inline]}

\definecolor{myblue}{rgb}{0.29, 0.59, 0.82}
\newcommand{\RevOne}[1]{{#1}}
\newcommand{\RevTwo}[1]{{#1}}
\newcommand{\RevThree}[1]{{#1}}
\newcommand{\Rev}[1]{{#1}}

\title{Rheology of mobile sediment beds in laminar shear flow: effects of creep and polydispersity}

\author{Christoph Rettinger\aff{1}
	\corresp{\email{christoph.rettinger@fau.de}},
	Sebastian Eibl\aff{1},
	Ulrich R{\"u}de\aff{1,2}
	\and Bernhard Vowinckel\aff{3}}

\affiliation{
	\aff{1}Chair for System Simulation, Friedrich--Alexander--Universit{\"a}t Erlangen--N{\"u}rnberg, Cauerstra\ss e 11, 91058 Erlangen, Germany
	\aff{2}CERFACS, 42 Avenue Gaspard Coriolis, 31057 Toulouse Cedex 1, France
	\aff{3}Leichtwei\ss-Institute for Hydraulic Engineering and Water Resources, Technische Universit\"at Braunschweig, 38106 Braunschweig, Germany}

\begin{document}
	
\maketitle

\begin{abstract}
Classical scaling relationships for rheological quantities such as the $\mu(J)$-rheology have become increasingly popular for closures of two-phase flow modeling. However, these frameworks have been derived for monodisperse particles. We aim to extend these considerations to sediment transport modeling by using a more realistic sediment composition. We investigate the rheological behavior of sheared sediment beds composed of polydisperse spherical particles in a laminar Couette-type shear flow. The sediment beds consist of particles with a diameter size ratio of up to ten, which corresponds to grains ranging from fine to coarse sand. The data was generated using fully coupled, grain resolved direct numerical simulations using a combined lattice Boltzmann - discrete element method. These highly-resolved data yield detailed depth-resolved profiles of the relevant physical quantities that determine the rheology, i.e., the local shear rate of the fluid, particle volume fraction, total shear, and granular pressure. A comparison against experimental data shows excellent agreement for the monodisperse case. We improve upon the parameterization of the $\mu(J)$-rheology by expressing its empirically derived parameters as a function of the maximum particle volume fraction. Furthermore, we extend these considerations by exploring the creeping regime for viscous numbers much lower than used by previous studies to calibrate these correlations. Considering the low viscous numbers of our data, we found that the friction coefficient governing the quasi-static state in the creeping regime tends to a finite value for vanishing shear, which decreases the critical friction coefficient by a factor of three for all cases investigated. 
\end{abstract}

\begin{keywords}
Done automatically
\end{keywords}


\section{Introduction}

The fluid mediated transport of granular sediment is a key process for the mass movement in a geophysical but also an engineering context \citep[e.g.][]{Frey2011}. The transport typically occurs along a slope or by a fluid flow shearing the sediment  \citep{jerolmack2019} and can lead to bedform evolution, such as ripples and dunes, even for laminar flow conditions \citep{lajeunesse2010a}. This consideration allows to characterize sediment transport in laminar flows in terms of the rheology to investigate the fluid-particle mixture's deformation behavior in shearing flows \citep{aussillous2013,houssais2016,Kidanemariam2017Diss,vowinckel2020}.
All these studies justified their approach by comparing the results to data previously obtained in rheometer studies with dense suspensions of neutrally buoyant particles \cite[e.g.][]{morris1999,boyer2011}. For these classical rheological investigations, a shear rate $\dot{\gamma}$ is applied to a dense granular material suspended in a fluid with viscosity $\eta_f$ to investigate the total shear stress $\tau$ acting on the fluid-particle mixture in the shearing direction and the imposed particle pressure $p_p$ in the wall-normal direction. The total shear comprises hydrodynamic and frictional inter-particle stresses, with the latter becoming more important with increasing particle volume fraction $\phi$ \citep{gallier2014,Guazzelli2018,vowinckel2020}. 

In this regard, two types of rheometer setups are possible. On the one hand, the volume-imposed rheometry confines the suspension by shearing walls with constant gap size \cite[e.g.][]{morris1999}. While \cite{morris1999} were investigating shear induced migration to begin with, they were also able to measure the effective shear and normal viscosities, $\eta_s=\tau/\eta_f\dot{\gamma}$ and $\eta_n=p_p/\eta_f\dot{\gamma}$, respectively, and to derive empirical correlations for these two quantities as functions of $\phi$. 
On the other hand, a pressure-imposed rheometer, where a constant confining pressure is applied to a movable upper wall, allows for the dilation of the dense suspension under shear \citep[e.g.][]{boyer2011,Dagois2015}. 
\Rev{
For laminar viscous flows, i.e. a Stokes number $St=\rho_p\dot{\gamma}d^2_p/\eta_f$ smaller than 10 \citep{bagnold1954,Ness2015}, where $\rho_p$ is the particle density and $d_p$ is the characteristic particle diameter, this measure allowed \cite{boyer2011} to define a macroscopic friction coefficient $\mu=\tau/p_p$ that depends on the viscous number $J=\eta_f\dot{\gamma}/p_p$. Based on this, the authors were able to propose empirical correlations for $\mu(J)$ and $\phi(J)$ that distinguish between stress contributions from particle contact and hydrodynamic interactions. This framework has become known as the $\mu(J)$-rheology. 
In this article, we will follow the nomenclature of \cite{Guazzelli2018} and use the symbol $J$ rather than $I_v$ for the viscous number to distinguish it more clearly from the inertial number defined for highly inertial granular flows.}

The pressure-imposed rheometry also allows for the analogy to sediment transport, where the imposed  particle pressure $p_p$ at some depth in the sediment bed is equal to  the submerged weight of the overlying grains \citep{aussillous2013,Maurin2016,vowinckel2019b}. This analogy is important for two-phase fluid sediment transport modeling \citep{jenkins1998,hsu2004}, where the fluid-particle mixture is treated as two separated continua with interconnected conservation laws of mass and momentum \citep{Ouriemi2009a}. The empirical correlations of the $\mu(J)$-rheology can provide the constitutive equations needed to close this set of equations \citep{chauchat2017,lee2018,lee2020}. Unfortunately, the empirical correlations $\mu(J)$ and $\phi(J)$ involve parameters that are not universal but were calibrated against the experimental data of \cite{boyer2011} in the dense regime with non-vanishing shear ($0.4<\phi<0.58$ and $J>10^{-6}$). It has been pointed out by \cite{Revil-Baudard2015} \Rev{who investigated sheet-flow processes under turbulent flow conditions } 
that these correlations need adjustments for more dilute systems, whereas \cite{houssais2016} investigated viscous numbers as low as $J\approx10^{-9}$ and found that the grains were still moving under creeping conditions even for these extremely low shear rates. It remained unclear, however, if this was a particle property or an effect originating from the curvature of the annual flume employed in this study. Hence, for cases, where the modeled flow conditions exceed the range of the calibration data, the $\mu(J)$-rheology can even lead to ill-posed problems as reported by \citet{barker2015}, who then proposed an extension to tackle this problem \citep{barker2017}. 

To increase the robustness of the $\mu(J)$-rheology for two-phase fluid models, more work is needed to derive more universal constitutive equations \citep{Denn2014,pahtz2019}. A good starting point will be to address the 
coefficients that enter the models of the $\mu(J)$-rheology and are known to depend on the particle properties. For the critical state of very low shear rates and dense systems, i.e. low $J$ and large $\phi$, the frictional inter-particle forces may become large enough to inhibit grains sliding past one another. This quasi-static regime is determined by the particle properties critical friction coefficient $\mu_1$
and maximum particle volume fraction $\phi_m$. For example, \cite{boyer2011} reported $\mu_1=0.32$ and $\phi_m=0.585$ for the monodisperse case, but it has been shown by \cite{tapia2019} for pressure-imposed rheometry that these two quantities decrease with increasing particle roughness. For obvious reasons, the critical volume fraction may also depend on the grain size distribution of the sediment as smaller particles can fill the void interstitial pore space provided in between larger grains \citep{Guazzelli2018}. This aspect has thus far been neglected in the framework of the $\mu(J)$-rheology. In fact, most of the studies use sediment compositions of uniform grains, where the standard deviation of the grain size distribution is smaller than 10\%. However, neither is this variance in grain size distribution large enough to see appreciable effects of polydispersity on the sediment transport \citep{biegert2017}, nor does this variance reflect the grain size distribution of fluvial sediments.

\Rev{In this regard, it is important to acknowledge that natural sediments are by no means monodisperse or bidisperse, but obey a certain continuous grain size distribution. For example, according to ISO 14688-1:2002, cohesionless sand grains can range from 0.063 to 2~millimeters in diameter. This calls for an extension of the $\mu(J)$-rheology towards more realistic polydisperse sediment compositions. 
}

As a first step, bidisperse suspensions were investigated in volume-imposed rheometers. For this scenario, the effective viscosities were reduced as compared to the monodisperse case \citep{chang1994,gondret1997}. In these studies, the non-uniformity of the bidisperse grains was up to $d_{p,\mathit{max}}/d_{p,\mathit{min}}=13.75$, where $d_{p,\mathit{max}}$ and $d_{p,\mathit{min}}$ are the maximum and minimum diameter of the grains, respectively. The critical volume fraction that indicates the quasi-static regime was also increased from $\phi_m=0.585$ for the monodisperse case \citep{boyer2011} to $\phi_m=0.64$. Consequently, models for $\phi_m$ in bi-disperse volume-imposed rheometry were proposed by \cite{dorr2013} and \cite{mwasame2016} that can also be applied to polydisperse systems \citep{pednekar2018}.

\Rev{As a next step, 2D-DEM simulations with grains of continuous polydispersity have been carried out where the fluid drag was approximated by Stokes drag and lubrication \citep{trulsson2012,Ness2015} and the variation of the grain size was kept constant at $d_{p,\mathit{max}}/d_{p,\mathit{min}}=3.0$ and $1.4$, respectively. 
A recent study by \cite{amarsid2017} extended these considerations to a lattice Boltzmann - discrete element method for simulations in 2D for $d_{p,max}/d_{p,min}=1.67$. 
Since, however, the focus of these studies was to investigate the transition from the viscous to the inertial regime, polydispersity was merely added to prevent artificial crystallization of the densely packed scenario and its role on the rheology was not discussed.
To the knowledge of the authors, 3D-simulations with a systematic focus on the degree of polydispersity in pressure-imposed rheometry or even sheared sediment beds have not been considered yet.
The present study addresses this issue.}

We employ the open-source simulation framework \walberla \, \citep{bauer2020} to carry out fully-coupled particle-resolved direct numerical simulations of sediment beds sheared by a laminar Couette-type flow in the viscous regime, i.e. $St<10$. 
To this end, we utilize the combined lattice Boltzmann - discrete element method of \cite{rettinger2017} and \cite{rettinger2020}.
This extends our pore-resolved simulations of fluid flow through porous media \citep{fattahi2016,gil2017,rybak2020}, and is in line with previous erosion studies using a similar methodology \citep{derksen2011,rettinger2017Riverbed}.
We follow the approach by \cite{vowinckel2020} to compute time-averaged, depth-resolved profiles to quantify the stress exchange between the fluid and the particle phase. This allows for a systematic simulation campaign of different sediment grain size compositions under exact control of the flow conditions and eradicates potentially unwanted  effects from curved sidewalls, as present in existing laboratory experiments. 
The highly-resolved data yields all the relevant quantities, i.e. particle volume fraction, shear rate, total shear, and granular pressure, to infer the rheology of the polydisperse fluid-particle mixture down to viscous numbers of $J\approx 10^{-9}$. The investigated sediment beds have a non-uniformity of $d_{p,\mathit{max}}/d_{p,\mathit{min}}$ up to a factor of ten, which corresponds to a variety typically encountered in fluvial sediments of lowland rivers \citep[e.g.][]{kuhnle1993,frings2008}. The rather large disparity of the grain sizes is achieved using the efficient parallelization scheme of \cite{eibl2018}. These studies ultimately allow us to derive a robust parameterization strategy of the classical $\mu(J)$-rheology to account for the sediment polydispersity by linking the non-uniformity to the critical volume fraction $\phi_m$ and propose a straightforward extension to creeping flow conditions that recovers the original $\mu(J)$-rheology for higher shear rates.

The paper is structured as follows. We first provide a brief summary of the numerical framework in \S\ref{sec:methods} and the simulation setup in \S\ref{sec:description}. We then infer the pressure-imposed rheology and validate our simulation approach in \S\ref{sec:rheology_monodisperse} by comparing the monodisperse case to the experimental data of \cite{boyer2011} and \cite{houssais2016}, including the classical empirical correlations of the $\mu(J)$-rheology \citep{boyer2011}. Finally, we utilize the data from our simulation campaign to present extensions of the $\mu(J)$-rheology for polydispersity and creeping flow in \S\ref{sec:polydisperse_extension} and \S\ref{sec:creep_extension}, respectively. 

\section{Numerical Method}\label{sec:methods}

\begin{figure}
	\centering
	\begin{tikzpicture}[]
		\fill[cyan!20!white] (0,0) rectangle (8.5,5);
		\fill[brown!30!white] (1,1.5) rectangle ++(6,1.5);
		\fill[brown!30!white] (1,1) rectangle ++(2.5,0.5);
		\fill[brown!30!white] (4,1) rectangle ++(3,0.5);
		\fill[brown!30!white] (1.5,0.5) rectangle ++(1,0.5);
		\fill[brown!30!white] (4.5,0.5) rectangle ++(2,0.5);
		\fill[brown!30!white] (1.5,3) rectangle ++(1.5,0.5);
		\fill[brown!30!white] (3.5,3) rectangle ++(3.5,0.5);
		\fill[brown!30!white] (3.5,3.5) rectangle ++(3.5,0.5);
		\fill[brown!30!white] (4,4) rectangle ++(2.5,0.5);
		\fill[brown!30!white] (7,2) rectangle ++(0.5,1.5);
		\draw[step=0.5,gray,very thin] (0,0) grid (8.5,5);
		
		
		\coordinate[label=left:$\boldsymbol{x}_{p,i}$] (xpi) at (2.2,2.1);
		\coordinate[label=right:$\boldsymbol{x}_{p,j}$] (xpj) at (5.4,2.6);
		\node[fill=black, circle, inner sep=1.5] at (xpi) {};
		\node[fill=black, circle, inner sep=1.5] at (xpj) {};
		\draw[black, <->] ($(xpi)+(0,-1.4)$) -- ++(0,2.8) node[pos=0.2,left]{$d_{p,i}$};
		\draw[black, <->] ($(xpj)+(0,-2)$) -- ++(0,4) node[pos=0.2,right]{$d_{p,j}$};
		\draw[orange,very thick] (xpi) circle (1.4);
		\draw[orange,very thick] (xpj) circle (2);
		
		\draw[-latex, thick] (xpi) -- ++(0.5,-0.5) node[pos=1,above right]{$\boldsymbol{u}_{p,i}$};
		\draw[-latex, thick] (xpj) -- ++(-0.4,0.9) node[pos=1,left]{$\boldsymbol{u}_{p,j}$};
		
		\draw[-latex,thick] ($(xpi) + (150:1.7)$) arc (150:95:1.7);
		\node[above] at ($(xpi) + (95:1.7)$) {$\boldsymbol{\omega}_{p,i}$};
		\draw[-latex,thick] ($(xpj) + (-10:2.3)$) arc (-10:35:2.3);
		\node[above] at ($(xpj) + (35:2.3)$) {$\boldsymbol{\omega}_{p,j}$};
		
		
	\end{tikzpicture}
	\caption{Schematic representation of the coupled LBM-DEM approach for fully-resolved particulate flow simulations. The orange circles depict two colliding spheres, $i$ and $j$. The underlying uniform grid is used for the LBM, which simulates the fluid flow inside the fluid (light blue) cells. The solid (light brown) cells, whose centers are contained inside the particles, do not carry fluid information.}
	\label{fig:numerical_method}
\end{figure}
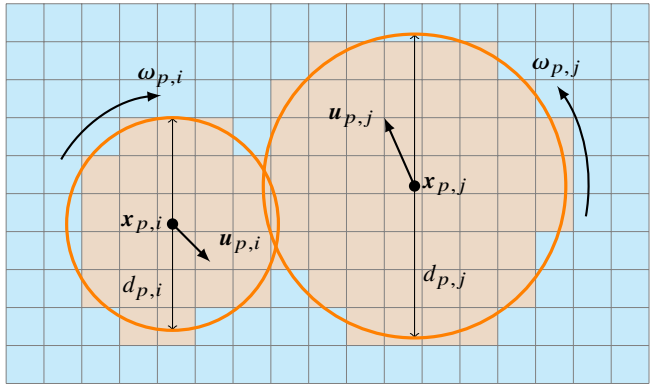

For the numerical studies presented here, we couple the lattice Boltzmann method for fluid flow with a discrete element method to account for particle interactions of polydisperse, spherical grains.
This approach has proven to be accurate and efficient for geometrically fully-resolved particle flow simulations and has been thoroughly validated in \cite{rettinger2020}.
Therein, a detailed presentation and discussion of the method is given.
We briefly summarize the key aspects for completeness here.
All parts of the employed numerical scheme are contained in the open-source high-performance framework \walberla{}~\citep[cf.][]{bauer2020}, and its implementation can be found in the official software repository\footnote{\url{https://walberla.net/}}.
A sketch of the numerical scheme is presented in figure \ref{fig:numerical_method}.

\subsection{Lattice Boltzmann method}

The lattice Boltzmann method (LBM) is a relatively recent approach for the simulation of viscous fluid flow. It describes the evolution of particle distribution functions (PDFs) on a uniform computational grid and thereby fulfills the macroscopic Navier-Stokes equations. A detailed overview of the theory and various approaches can be found in \cite{krueger2017}.
For the present studies, we employ the $D3Q19$ two-relaxation-time model of \cite{ginzburg2008}.
The relaxation times, connected via the parameter $\Lambda=3/16$, determine the kinematic fluid viscosity $\nu_f$ and allow for accurate flow simulations.
The local fluid pressure $p_f$ and velocity $\boldsymbol{u}_f$ are obtained via zeroth- and first-order moments of the PDFs in a fluid cell.  
Commonly, all quantities are expressed in a normalized LBM unit system, the so-called lattice units, which results in the cell size $\Delta x = 1$, the time step size $\Delta t = 1$, and a reference fluid density of $\rho_f = 1$.
Those will be used in the remainder of this work.

\subsection{Discrete element method}\label{sec:DEM}

The motion of a spherical particle $i$ can be described by the Newton-Euler equations
\begin{align}
	m_{p,i} \frac{\text{d} \boldsymbol{u}_{p,i}}{\text{d} t} &=  \boldsymbol{F}_{p,i} = \boldsymbol{F}_{p,i}^\mathit{col} + \boldsymbol{F}_{p,i}^\mathit{hyd} + \boldsymbol{F}_{p,i}^\mathit{ext}, \\
	I_{p,i} \frac{\text{d} \boldsymbol{\omega}_{p,i}}{\text{d} t} &= \boldsymbol{T}_{p,i} = \boldsymbol{T}_{p,i}^\mathit{col} + \boldsymbol{T}_{p,i}^\mathit{hyd}.
\end{align}
Here, $m_{p,i} = \rho_{p} V_{p,i}$ is the mass of the particle of density $\rho_{p}$ and volume $V_{p,i}$, and ${I_{p,i} = (m_{p,i} d_{p,i}^2)/10}$ is the moment of inertia for a sphere of diameter $d_{p,i}$.
The temporal change of the particle's translational velocity is thus given by the acting forces $\boldsymbol{F}_{p,i}$, with contributions from the collisions $\boldsymbol{F}_{p,i}^\mathit{col}$, the hydrodynamic interactions $\boldsymbol{F}_{p,i}^\mathit{hyd}$ and external sources $\boldsymbol{F}_{p,i}^\mathit{ext}$.
Similarly, the angular velocity changes according to the acting torque $\boldsymbol{T}_{p,i}$, due to collisions and hydrodynamic interactions.
These equations, together with the particle's position, are integrated in time via a Velocity Verlet scheme \citep{wachs2019} with a constant time step size $\Delta t_p = \Delta t/10$. 
Consequently, ten particle simulation time steps are carried out within one fluid time step, which improves the overall accuracy of particle interactions and the efficiency of the simulation. 

The collision forces and torques are determined via a discrete element method (DEM) that assumes a soft contact between overlapping rigid particles  \citep[cf. ][]{cundall1979}.
In our case, the normal and tangential collision components are given by a linear spring-dashpot model, similar to \cite{costa2015} and \cite{biegert2017}.
Following \cite{vanDerHoef2006}, the spring and damping coefficients of the normal collision model, $k_n$ and $d_n$, are determined via the dry coefficient of restitution $e_\mathit{dry}$, a material parameter that is here chosen to be $0.97$ \citep{vowinckel2020}, and the collision time $T_c$.
The latter is chosen according to the findings in \cite{rettinger2020} as
$T_c = 4 \bar{d}_p \Delta t /\Delta x$, where $\bar{d}_p$ is an average particle diameter, and ensures an adequate temporal resolution of the collision.
As shown in \cite{thornton2013}, the spring and damping coefficient of the tangential model are related to the ones of the normal direction via the Poisson's ratio $\nu_p$, such that $k_t = \kappa_p k_n$ and $d_t = \sqrt{\kappa_p}d_n$, with $\kappa_p = 2(1-\nu_p)/(2-\nu_p)$.  
The magnitude of the tangential collision force is limited by the Coulomb friction, determined as a product of the friction coefficient $\mu_p$ and the absolute value of the normal collision force.
In the present simulations, we use $\nu_p = 0.22$ and $\mu_p=0.15$ as reported in \cite{joseph2004}.

The external force is given as the gravitational and buoyancy forces due to the gravitational acceleration $\boldsymbol{g}$, i.e. $\boldsymbol{F}_{p,i}^\mathit{ext} = (\rho_p-\rho_f) V_{p,i} \boldsymbol{g}$.

\subsection{Fluid-particle coupling}

To establish the coupling between the fluid and the granular phase in an accurate manner, we follow \cite{rettinger2020} and distinguish between resolved and unresolved hydrodynamic forces to compute $\boldsymbol{F}_{p,i}^\mathit{hyd}$ and $\boldsymbol{T}_{p,i}^\mathit{hyd}$. For the resolved part, we use the LBM-specific momentum exchange method as proposed by \cite{aidun1998} to apply an explicit mapping of the particles onto the computational grid.
This is achieved by flagging cells with their centers contained inside of particles as solid, effectively removing them from the fluid domain (cf. figure \ref{fig:numerical_method}).
This results in a sharp interface between the fluid and solid phase, along which no-slip boundary conditions for the fluid are applied.
Here, we use the central linear interpolation (CLI) scheme of \cite{ginzburg2008} that allows for second-order accurate results by including information about the exact surface position.
The momentum exchanged locally with the particle due to its no-slip boundary condition is then integrated over the whole particle surface, as in \cite{wen2014}.
Following \cite{ladd1994}, this measure determines the resolved part of the fluid-particle interaction force $\boldsymbol{F}_{p,i}^\mathit{fp}$ and torque $\boldsymbol{T}_{p,i}^\mathit{fp}$ acting on this particle,
which are averaged over two consecutive fluid time steps for improved stability.
Solid cells that are no longer occupied by the particle due to its motion are converted back to fluid cells.
Additionally, the otherwise missing PDF information is restored in these cells with an approach similar to \cite{dorschner2015}, using density and pressure tensor information from surrounding fluid cells and the particle's velocity.

As shown in \cite{rettinger2020}, this approach is able to reliably and accurately predict the resolved part of the fluid-particle interactions of single spheres.
For two approaching particles, however, the mesh resolution of the narrow gap between the particles' surfaces is usually too coarse to fully resolve the strong lubrication interaction originating from the fluid that is being squeezed out of the gap of size $\delta_n$.
For those cases, a lubrication correction model must be applied that accounts for these unresolved forces \citep{nguyen2002,biegert2017}.
Thus, the total hydrodynamic interaction force and torque on a particle $i$ is here computed as
\begin{align}
	\boldsymbol{F}_{p,i}^\mathit{hyd} &= \boldsymbol{F}_{p,i}^\mathit{fp} + \boldsymbol{F}_{p,i}^\mathit{lub,cor}, \label{eq:hydrodynamic_force}\\
	\boldsymbol{T}_{p,i}^\mathit{hyd} &= \boldsymbol{T}_{p,i}^\mathit{fp} +  \boldsymbol{T}_{p,i}^\mathit{lub,cor}. \label{eq:hydrodynamic_torque}
\end{align}
These lubrication correction forces and torques explicitly account for the pair-wise lubrication forces and torques due to relative normal, tangential translational, and tangential rotational velocities, and are given in \cite{rettinger2020}.
As suggested by validation studies therein, the normal and tangential lubrication corrections are only active for $\delta_n<2\Delta x/3$ and $\delta_n<\Delta x/2$, respectively.
As these corrections scale as $\boldsymbol{F}_{p,i}^\mathit{lub,cor} \propto \delta_n^{-1}$ and $\boldsymbol{T}_{p,i}^\mathit{lub,cor} \propto \ln(\delta_n)$, they would grow to infinity for vanishing gap sizes. Hence, a calibrated lower limit of ${\delta_{n,\mathit{min}}^\mathit{lub}=(0.001 + 0.000035 d_{p,i}/\Delta x)\,d_{p,i}/2}$ is applied in their calculation.

\section{Simulation description}\label{sec:description}

In this section, we detail the set up of the simulation, including the generation of the sediment beds, the physical parameterization, the description of the computational setup, and, finally, the evaluation of relevant rheological quantities.
\subsection{Setup description}\label{sec:physical_parameterization}

\begin{figure}
\centering
\begin{tikzpicture}
\node[inner sep=-1pt] (setup) at (0,0)
{\includegraphics[width=.7\textwidth]{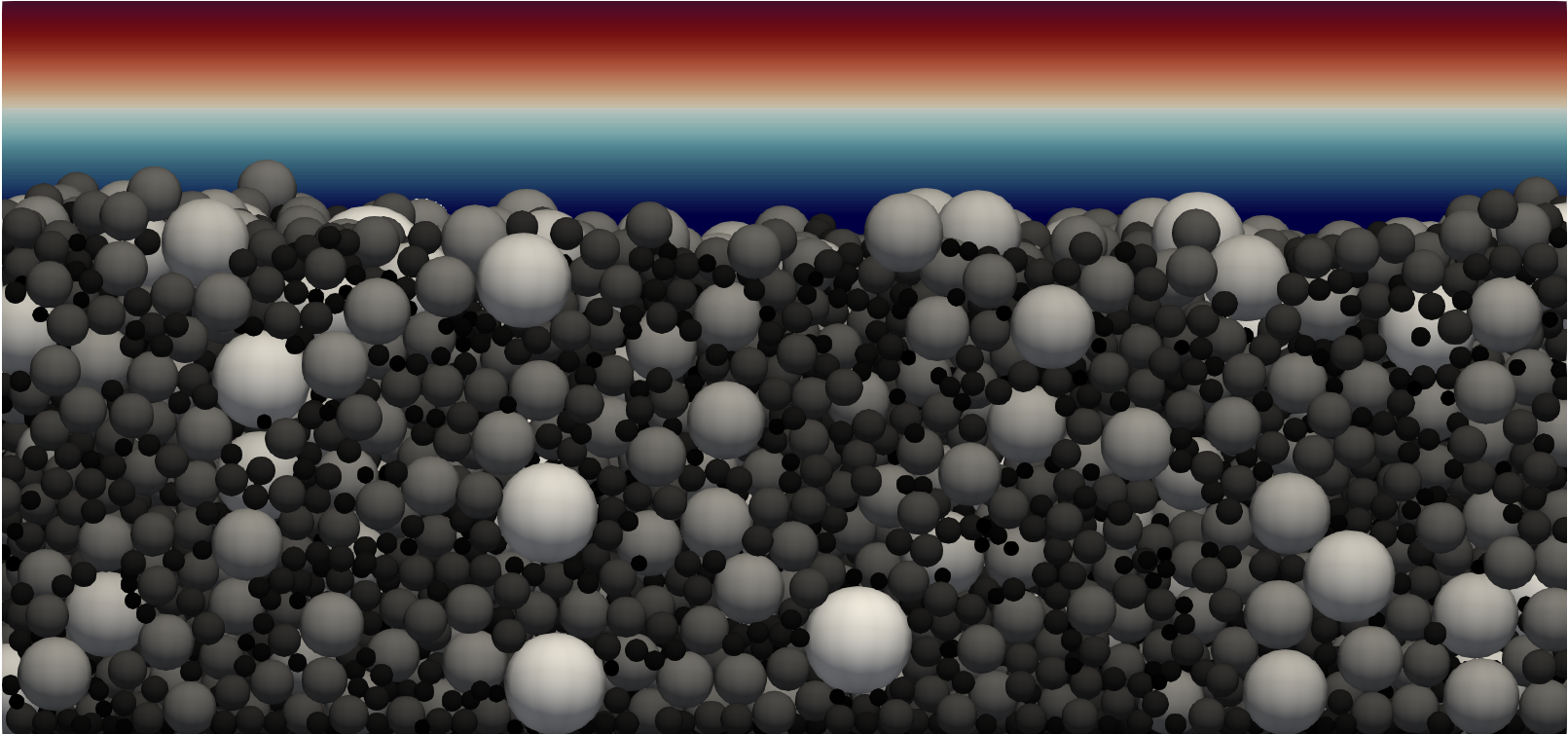}};
\draw[<->,thick] ([yshift=-1ex]setup.south west) -- ([yshift=-1ex]setup.south east) node[midway,anchor=north] {$L_x$};
\draw[<->,thick] ([xshift=-1ex]setup.south west) -- ([xshift=-1ex]setup.north west) node[midway,anchor=east] {$L_z$};
\coordinate (A) at ($(setup.south west)!0.71!(setup.north west)$);
\coordinate (B) at ($(setup.south east)!0.71!(setup.north east)$);
\draw[-,very thick,orange] (A) to (B);
\draw[<->,thick] ([xshift=1ex]setup.south east) -- ([xshift=1ex]B) node[midway,anchor=west] {$h_b$};
\draw[<->,thick] ([xshift=1ex]setup.north east) -- ([xshift=1ex]B) node[midway,anchor=west] {$h_f$};

\coordinate (C) at (setup.north);
\coordinate (D) at ($(A)!0.5!(B)$);
\coordinate (E) at ($(C)+(1,0)$);

\draw[thick,white] (C) -- (D);
\draw[thick,white] (D) -- (E);
\draw[-latex,thick,white] ([yshift=-1pt]C) -- ([yshift=-1pt]E) node[midway,anchor=north]{$U_w$};

\draw[-latex,thick] ([xshift=8ex]setup.east) --++(0,-1) node[midway,anchor=west]{$\boldsymbol{g}$};
\end{tikzpicture}
\caption{Sketch of physical setup as a side-view, including a slice of the initial flow field above the sediment bed.}
\label{fig:sketch_setup}
\end{figure}
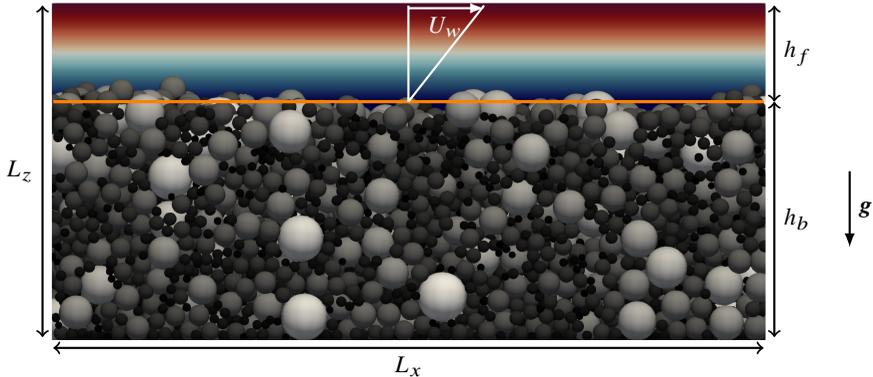

The general scenario is to consider linear shear flows with a constant shear rate $\dot{\gamma}=U_w/h_f$ across sediment beds of polydisperse, spherical particles (cf. figure~\ref{fig:sketch_setup}), where $U_w$ is the velocity of the moving top wall, $h_f=L_z-h_b$ is the clear fluid height, $L_z$ is the vertical extent of the domain, and $h_b$ is the height of the sediment. To this end, we generate a grain size distribution with diameter values for $N_p$ particles by sampling from a log-normal distribution, defined by the parameters $\mu_\mathit{LN}$ and $\sigma_\mathit{LN}^2$. Those parameters are related to the desired mean $\mu_X$ and variance $\sigma_X^2$ of the distribution via
\begin{equation}
\mu_\mathit{LN} = \ln\left(\frac{\mu_X^2}{\sqrt{\mu_X^2 + \sigma_X^2}}\right) \text{ and } \sigma_\mathit{LN}^2 = \ln\left(1+ \frac{\sigma_X^2}{\mu_X^2}\right),
\end{equation}
which yields the mean diameter
\begin{equation}
\bar{d}_p = \frac{1}{N_p} \sum_{i=1}^{N_p} d_{p,i}.
\label{eq:average_diameter}
\end{equation}
Note, that we decided to use the arithmetic mean diameter for the parameterization instead of the median diameter $d_{p,50}$ as it is also well-defined for bidisperse grain size distributions. As will be detailed in \S \ref{sec:parameterization}, we target a numerical resolution of the mean diameter of $\bar{d}_p/\Delta x=20$. Especially for large variances, care must be taken to maintain a reasonable numerical resolution for all particle sizes including its smallest values. Hence, we dismiss diameter values below $10$ cells to guarantee a reasonable resolution of the flow field around the particles.

\begin{table}
	\centering
	\begin{tabular}{lccccccc}
	\toprule
		case & $N_p$ & $\mu_X$ & $\sigma_X^2$  & $\bar{d}_p$ & $d_{p,\mathit{max}}/d_{p,\mathit{min}}$ & $d_{p,50}$ & $h_b^0 / \bar{d}_p$ \\ \midrule
		\mono & 26112 & 20 & 0.1 & 20.00 & 1.15 & 20.00 & 17.10 \\
		\polya & 24486 & 20 & 10 & 20.02 & 3.43 & 19.78 & 17.00 \\
		\polyb & 19404 & 19.5 & 50 & 20.00 & 7.87 & 18.72 & 17.02 \\
		\polyc & 14464 & 17.5 & 100  & 20.27 & 9.74 & 17.60 & 16.77\\
	\bottomrule
	\end{tabular}
	\caption{Parameters and properties of the different sediment beds, where length scales are expressed in lattice units.}
	\label{tab:bed_properties}
\end{table}

The statistical properties of the polydisperse sediments including the ratio of largest to smallest diameter in the bed, given by $d_{p,\mathit{max}} = \max_i d_{p,i}$ and $d_{p,\mathit{min}} = \min_i d_{p,i}$, can be found in table~\ref{tab:bed_properties}. Note that $\mu_X$ was chosen below $20$ for strong polydispersity to compensate for the lower limit of admissible diameters and to obtain $\bar{d}_p / \Delta x \approx 20$. We also use a log-normal distribution, albeit with a much smaller variance, for the monodisperse case as encountered in experimental studies~\citep{boyer2011,aussillous2013} to prevent an artificially close packing observable in perfectly mono-sized sphere beds.

\begin{figure}
\begin{subfigure}{0.5\textwidth}
	\centering
	\includegraphics[trim=50 130 50 280, clip, width=\textwidth]{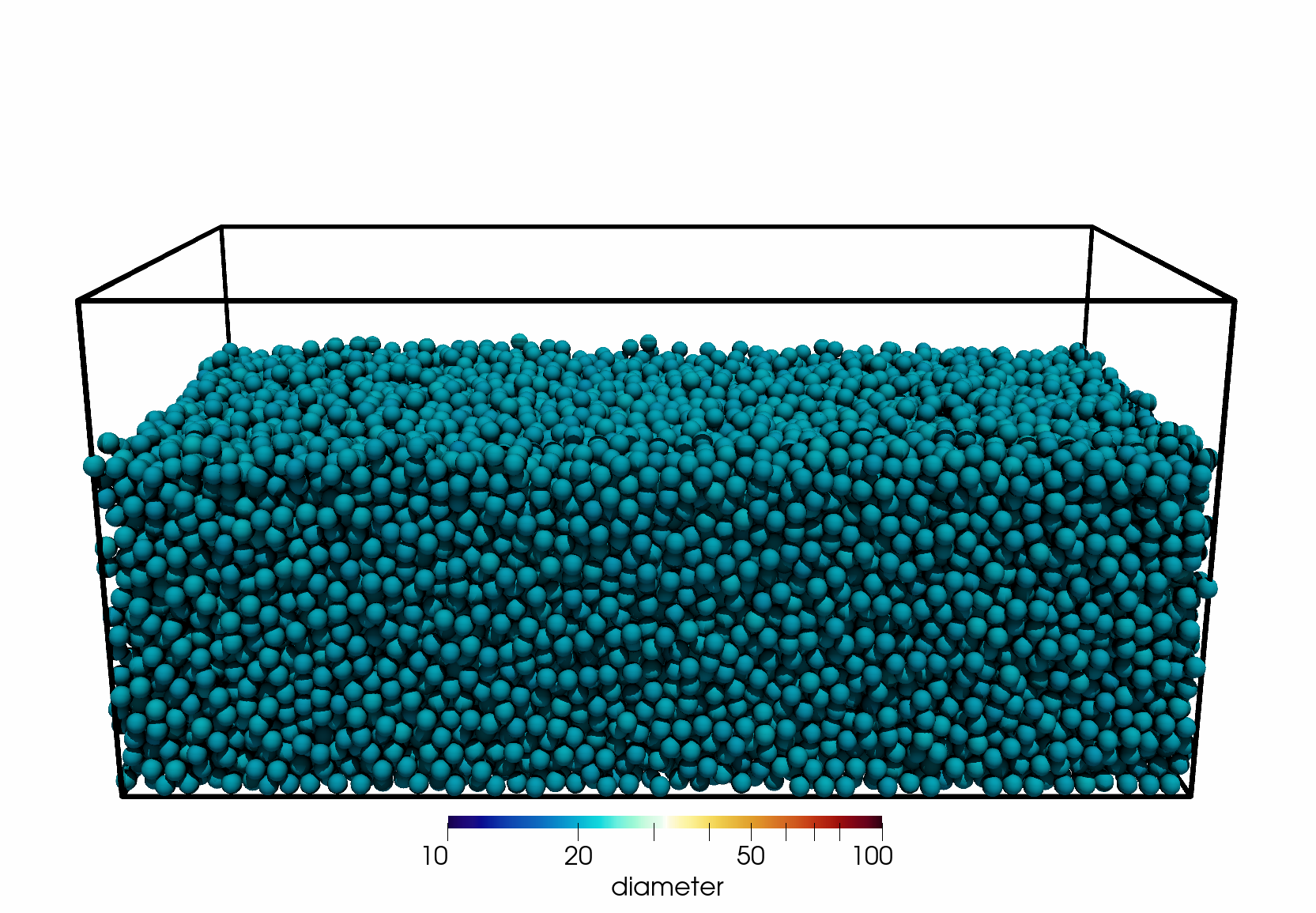}
\end{subfigure}%
\begin{subfigure}{0.5\textwidth}
	\begin{flushright}
	\input{figures/diameter_distribution_var01.pgf}
	\end{flushright}
\end{subfigure}

\begin{subfigure}{0.5\textwidth}
	\centering
	\includegraphics[trim=50 130 50 280, clip, width=\textwidth]{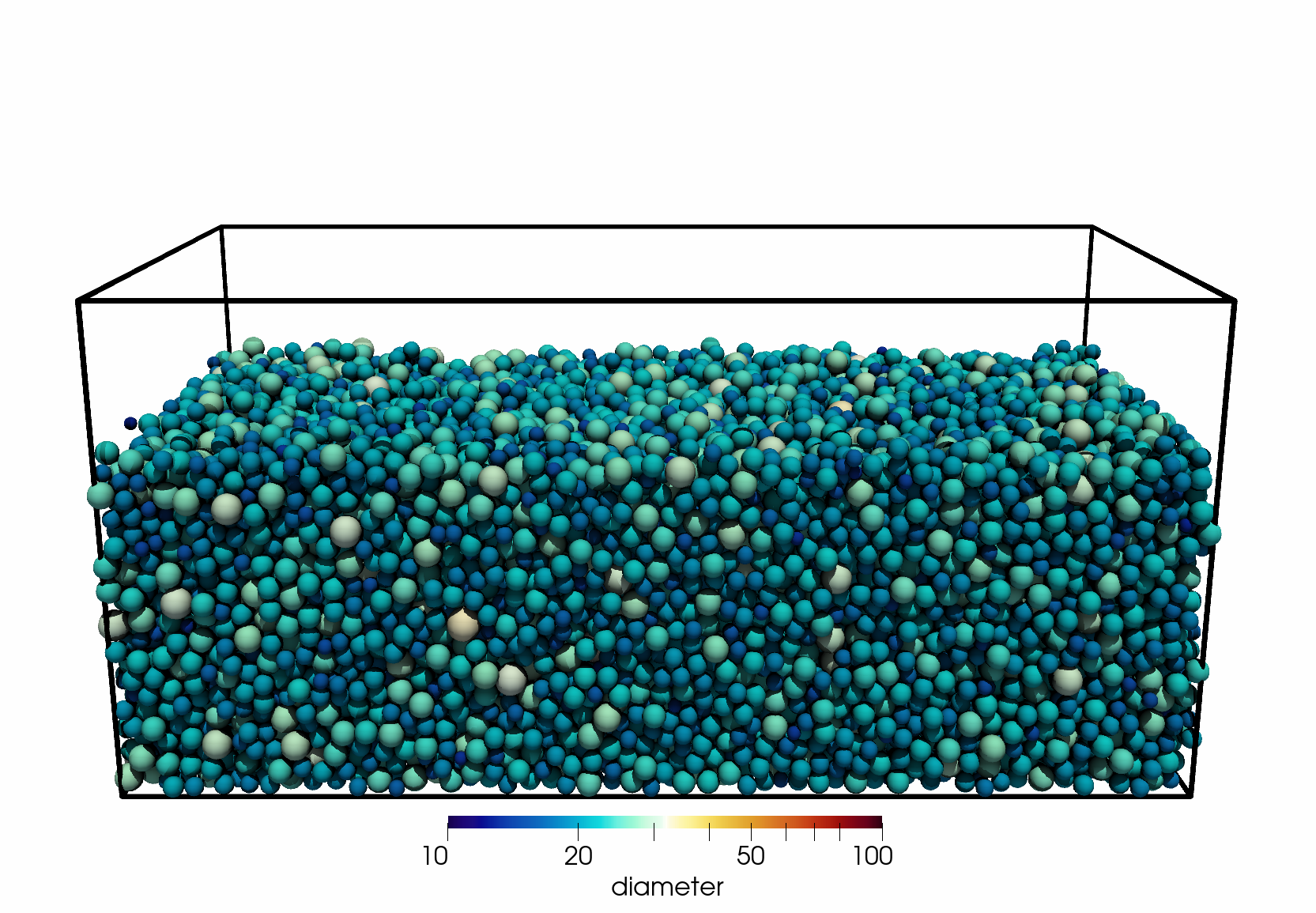}
\end{subfigure}%
\begin{subfigure}{0.5\textwidth}
	\begin{flushright}
	\input{figures/diameter_distribution_var10.pgf}
	\end{flushright}
\end{subfigure}

\begin{subfigure}{0.5\textwidth}
	\centering
	\includegraphics[trim=50 130 50 280, clip, width=\textwidth]{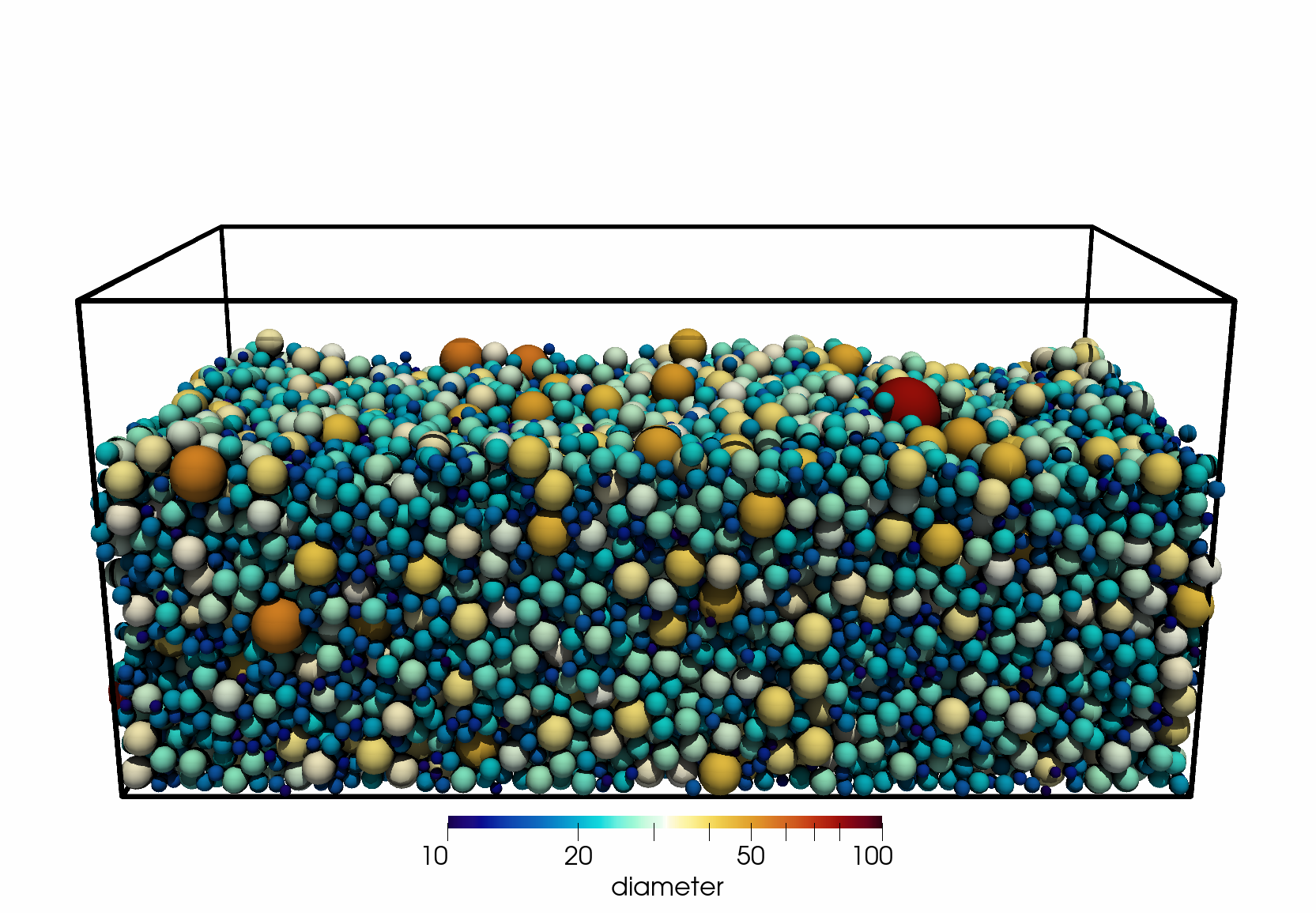}
\end{subfigure}%
\begin{subfigure}{0.5\textwidth}
	\begin{flushright}
	\input{figures/diameter_distribution_var50.pgf}
	\end{flushright}
\end{subfigure}

\begin{subfigure}{0.5\textwidth}
	\centering
	\includegraphics[trim=50 0 50 280, clip, width=\textwidth]{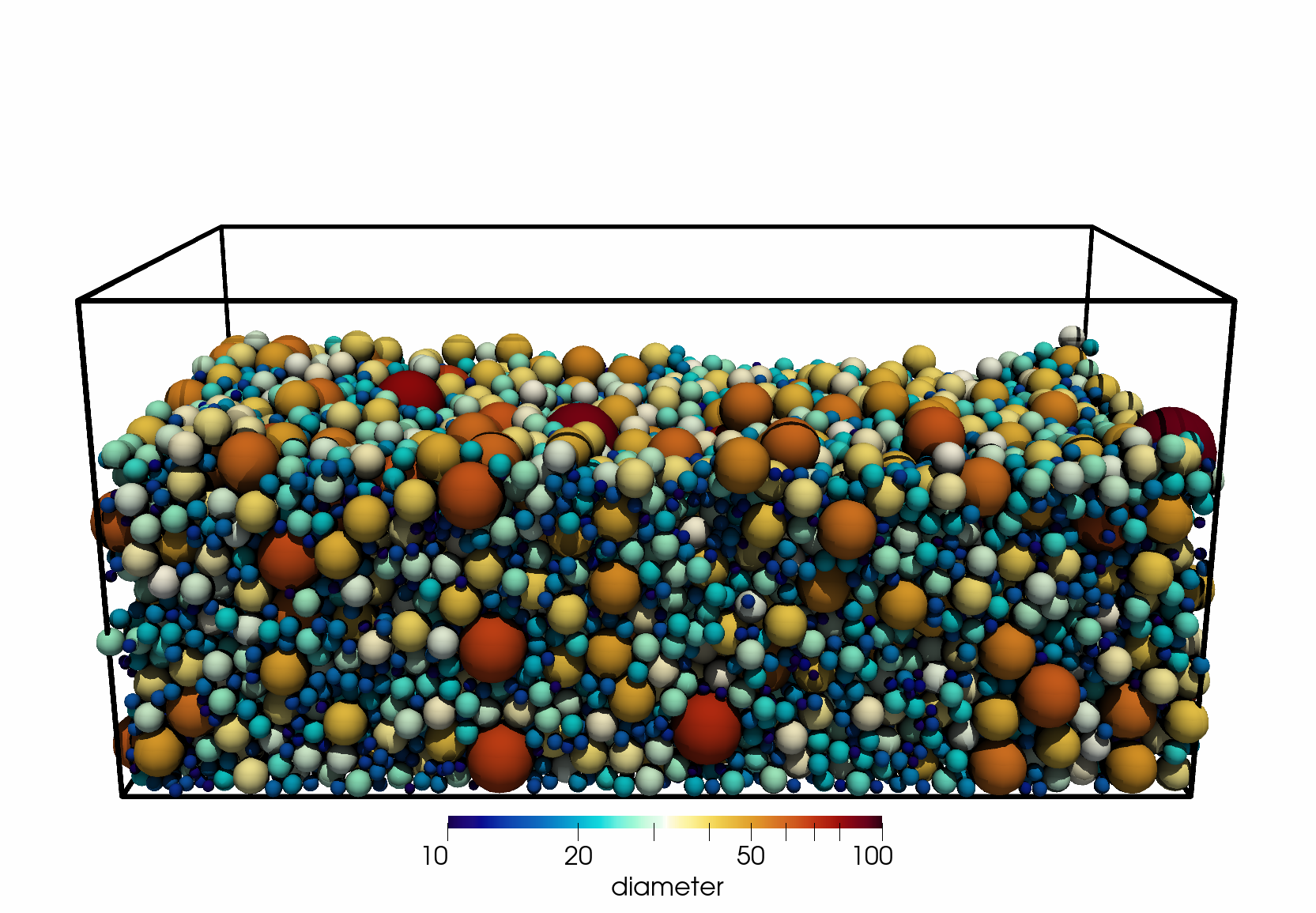}
\end{subfigure}%
\begin{subfigure}{0.5\textwidth}
	\begin{flushright}
	\input{figures/diameter_distribution_var100.pgf}
	\end{flushright}
\end{subfigure}
\caption{The four different setups and their diameter distribution (from top to bottom: \mono, \polya, \polyb, \polyc). Coloring of particles is according to the diameter with a logarithmic color scale. See table \ref{tab:bed_properties} for detailed information about bed configurations. Along the diameter distribution, the cumulative distribution function (CDF) based on a kernel density estimate is provided.}
\label{fig:bed_visualization}
\end{figure}

Subsequently, the initial sediment beds for the main simulations of a fully coupled fluid-particle system are created by a precursor simulation without fluid. A constant density $\rho_p$ is assigned to the particles. 
Initially, they are placed inside a tall domain, with a uniform spacing in all directions that prevents potentially large overlaps, and given a random velocity.
Due to gravity, they then settle due to gravity on a plate of size $L_x\times L_y=51.2\bar{d}_p \times 25.6 \bar{d}_p = 1024\times 512$ cells, where $L_x$ and $L_y$ are the streamwise and spanwise extent, respectively, of the horizontally periodic computational domain. The precursor simulations are run until all particles have come to rest to yield the initial bed height $h_b^0$ for the main simulation. This state is typically achieved after some minutes of simulation time on a regular workstation.  We noticed that this precursor simulation requires the same physical parameters, such as gravitational acceleration and submerged weight, as in the main simulation to prevent large accelerations followed by abrupt position changes in the initial phase of the main simulation. Since $h_b^0$ can only be roughly estimated {\em a priori}, an iterative procedure is applied to find the right amount of particles $N_p$ necessary to achieve comparable bed heights among the different runs.
In all cases, the bed is generated to obtain an initial bed height of approximately $h_b^0 = 340$, i.e. $h_b^0 / \bar{d}_p = 17$ (cf. table~\ref{tab:bed_properties}). 
This requires around 26000 particles for the monodisperse case to around 14500 particles for the strongly polydisperse setup.
A visualization of the generated sediment beds and the diameter distribution for all four cases can be seen in figure \ref{fig:bed_visualization}.

\subsection{Physical parameterization}\label{sec:parameterization}

The main simulation is executed in a cuboidal domain of size $L_x\times L_y \times L_z = 1024 \times 512 \times 480$ cells. The domain is completely filled with a viscous fluid, defined by the kinematic viscosity $\nu_f$ and density $\rho_f$. Periodic boundary conditions are applied in streamwise ($x$) and spanwise ($y$) direction, while no-slip boundaries are applied at the particle surface as well as the top and bottom planes bounding the vertical direction ($z$). The top plane is moving in $x$-direction with a constant velocity $U_w = 0.03$ in lattice units. The sphere packing is initialized by the results from the precursor simulations to prescribe $h_{b}^0$. We fix all particles with a vertical center position smaller than $3/4\bar{d}_p$ throughout the simulation to form a bottom roughness. This measure prevents artificial slipping of the complete bed over the bottom plane~\citep{jain2017,biegert2017}. A linear shear profile is assigned to the fluid above the sediment bed as an initial condition (cf. figure~\ref{fig:bed_visualization}).

Apart from the density ratio $\rho_p/\rho_f$, we characterize the sediment mobility by the Shields parameter $\Theta$:
\begin{equation}
\Theta = \frac{\tau}{g (\rho_p-\rho_f) \bar{d}_p},
\label{eq:shields}
\end{equation}
where $\tau  = \rho_f \nu_f \dot{\gamma}$ is the shear stress, and $g$ is the magnitude of the gravitational acceleration. Additionally, we define a particle Reynolds number $Re_p = u_\tau \bar{d}_p/\nu_f$ using $u_\tau = \sqrt{\tau /\rho_f}$.

For those non-dimensional parameters, we choose $\Theta = 0.5$, $Re_p = 0.76$, and $\rho_p/\rho_f = 1.5$ in all simulations to have comparable results. 
The value of the Shields parameter is well above the expected threshold for incipient motion, given as $\Theta_c \approx 0.12$ by \cite{ouriemi2007}, to ensure an adequate mobility of the particles. 
This results in a bulk Reynolds number based on channel properties $Re_b = U_w h_f/(2 \nu_f)$ of around $14$ and a Stokes number, $St = \rho_p \bar{d}_p^2 \dot{\gamma}/\eta_f$, of around $0.85$, which makes the simulations fall into the viscous regime \citep{bagnold1954}. Due to the low Reynolds number, we obtain a laminar Couette-like flow profile in the bulk region above the bed, where $\tau$ is constant.
Finally, we define the reference time scale as $t_\mathit{ref} = \bar{d}_p / U_w$. We explicitly note that the set of physical parameters of the simulations is determined using the initial values of the bed and the fluid height, since $h_b$ becomes a result of the simulation and varies over time when the sediment bed dilates under shear as will be detailed in \S\ref{sec:evaluation}.

To ensure an accurate resolution of fluid-particle interaction, a numerical resolution of approximately $20$ cells per mean diameter is chosen in all simulations, i.e. $\bar{d}_p/\Delta x \approx 20$ \citep{rettinger2017,costa2015,biegert2017,rettinger2020}. Since such a high resolution  inherently renders the present numerical simulations computationally challenging, a performance-optimized implementation of the numerical methods as well as efficient communication routines must be applied to stay within adequate runtimes without exhausting computational resources \citep{eibl2018,bauer2020_lbmpy}. The details of our simulation approach are presented in \cite{bauer2020}. The approach has successfully been applied in previous large-scale studies of particle-resolved simulations \citep[e.g.][]{goetz2010,rettinger2017Riverbed}, where its excellent performance on HPC-clusters has been demonstrated. Specifically in the present work, each simulation run is executed for 48h on $7680$ processes on the SuperMUC-NG supercomputer at LRZ in Garching, Germany.
The resulting $2.5\times10^8$ grid cells, simulated for around $9\times 10^6$ time steps in each case, make the studies at hand one of the largest and computationally most costly simulation campaigns of polydisperse sediment beds reported in literature.

Movies of the simulations are provided as supplementary material.

\subsection{Evaluation procedure for simulation data}\label{sec:evaluation}

\begin{figure}
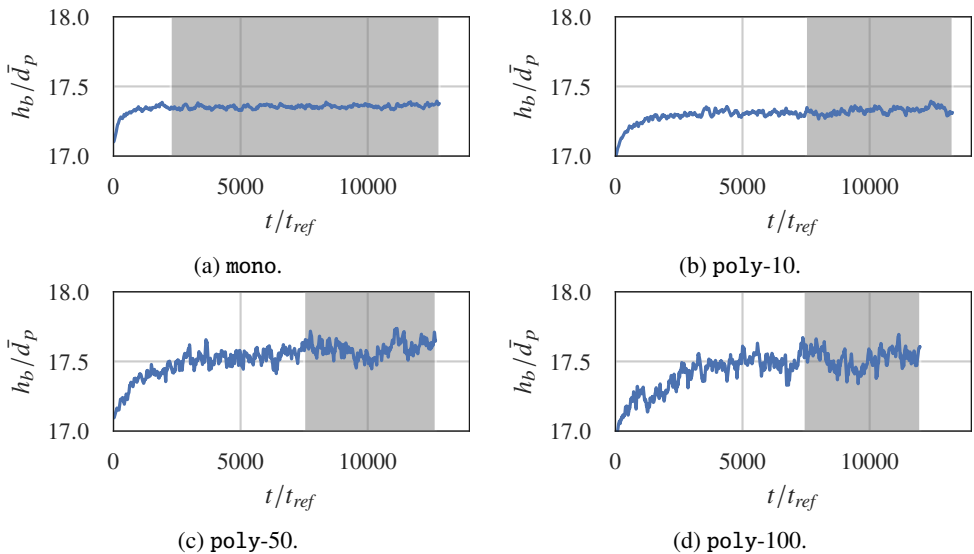

	\begin{subfigure}{0.48\textwidth}
		\centering
		\input{figures/bedHeight_over_time_var01.pgf}
		\caption{\mono.}
	\end{subfigure}~
	\begin{subfigure}{0.48\textwidth}
		\centering
		\input{figures/bedHeight_over_time_var10.pgf}
		\caption{\polya.}
	\end{subfigure}

	\begin{subfigure}{0.48\textwidth}
		\centering
		\input{figures/bedHeight_over_time_var50.pgf}
		\caption{\polyb.}
	\end{subfigure}~
	\begin{subfigure}{0.48\textwidth}
		\centering
		\input{figures/bedHeight_over_time_var100.pgf}
		\caption{\polyc.}
	\end{subfigure}
	\caption{Bed height $h_b$ as a function of time extracted from the instantaneous vertical volume fraction profiles for all simulation setups. The gray area depicts the region used for temporal averaging.}
	\label{fig:bed_quantities_over_time}
\end{figure}

Since the goal of the present study is to investigate the rheological behavior of sediment beds in the framework of the $\mu(J)$-rheology, we have to obtain the values for $p_p$, $\mu=\tau/p_p$, and $J=\eta_f \dot{\gamma}/p_p$. These quantities can be determined from vertical profiles of $\dot{\gamma}$ and $\phi$ \citep{houssais2016,vowinckel2020}. From the numerical simulations, we obtain high fidelity data of individual particle positions and velocities, as well as flow velocities as a function of time and space. To process the data for robust rheological interpretations, we apply spatial and temporal averaging.

As a first step, we perform spatial averaging and analyze it over time to determine the initialization period needed to obtain a statistically stationary state. 
\Rev{This measure ensures that transient effects such as the dilation of the granular packing under shear and the initial sorting of the polydisperse grains are excluded from the statistical analysis (cf. Appendix \ref{sec:app1}).}
We subdivide the domain into binned averaging volumes of size $V_0=L_x \times L_y \times \Delta x$, stacked vertically upon each other. 
In order to obtain the vertical particle volume fraction profile at a specific time $t$, we make use of the particle diameter and its center coordinates. 
The horizontal planes between the stacked $V_0$ slice each sphere into several sphere segments, whose volume $V_s$ can be determined analytically. 
We then add up all the volumes of the sphere segments within a $V_0$ and divide this accumulated particle volume $\sum V_s$ by the total averaging volume $|V_0|$ 
to obtain the particle volume fraction $\phi(z,t)$, where $z$ the discrete vertical center coordinate of the respective $V_0$. 
As a next step, we apply a central moving average of width $10\Delta x$, which corresponds to half of the mean particle diameter. 
This measure is needed to even out the layering at the sub-particle scale that introduces fluctuations within horizontally averaged profiles \citep{vowinckel2020}.

From these vertical profiles and with linear interpolation, we can evaluate the bed height $h_b(t)$ given as the vertical position, for which $\phi(h_b,t) = 0.1$ ~\citep{kidanemariam2014}. 
Note that other authors have used different threshold values for this definition~\citep{houssais2016,biegert2017}, but due to the sharp gradient of the profile at the interface region, the actual value to determine $h_b$ does not have an impact on our analysis of rheological quantities. The temporal evolution of the bed height due to the movement of the top particle layer is illustrated in figure~\ref{fig:bed_quantities_over_time}. It can be seen that when increasing the polydispersity of the bed, fluctuations in $h_b$ become larger and also, on average, the bed expands more.

Based on these evaluations, we define an instant of time that marks the beginning of our averaging time, $t_0$. 
As mentioned above, this is done to exclude the initial dilation phase of the sediment bed and, in particular, \RevOne{possible morphological effects due to vertical grain size segregation for the polydisperse cases, see Appendix \ref{sec:app1}}. 
\RevTwo{Hence, no significant changes in the rheological quantities nor the local particle size distributions are observed during the evaluation period.}
The temporal averaging windows for the different cases are stated in table~\ref{tab:bed_properties} and visualized in figure~\ref{fig:bed_quantities_over_time} as gray shaded areas. The slightly different end times originate from the different total run time of the simulations.

These considerations finally allow us to obtain the time-averaged particle volume fraction as
\begin{equation}
\langle \phi \rangle_{t}(z) = \frac{1}{t_1-t_0} \int_{t_0}^{t_1} \phi(z,t) \text{\,d}t, \label{eq:temporal_averaging}
\end{equation}
where the angular brackets indicate averaging in time as implied by the subscript $t$.
Similarly, we evaluate the time-averaged bed height and state it in table~\ref{tab:setups_evaluation}.

\begin{table}
	\centering
	\begin{tabular}{lcccc}
	\toprule
		case & $[t_0,t_1] / t_\mathit{ref}$ & $\langle h_b\rangle_t/\bar{d}_p$ & $\tau / \left(g (\rho_p-\rho_f) \bar{d}_p\right) $ & $\phi_m$\\\midrule
		\mono  & $[2250,12824]$ & $17.36$  & $0.506$ & $0.631$ \\
		\polya & $[7494,13264]$ & $17.33$  & $0.502$ & $0.645$ \\
		\polyb & $[7500,12674]$ & $17.59$  & $0.519$ & $0.669$ \\
		\polyc & $[7402,11991]$ & $17.52$  & $0.526$ & $0.697$ \\
		\bottomrule
	\end{tabular}
	\caption{Sediment bed and flow quantities extracted from the simulation data, together with duration of the time-averaging period.}
	\label{tab:setups_evaluation}
\end{table}

Analogously, we perform the spatial and temporal averaging of the streamwise fluid velocity $u_f$.
There, we define an indicator function $\Gamma$ being 1 in the fluid and 0 otherwise that separates the fluid from the particle phase to compute so-called intrinsic spatial averages \citep{vowinckel2017,vowinckel2019}:
\begin{equation}
\langle u_f \rangle_{V}(z,t) = \frac{1}{\int_{V_0} \Gamma  \text{\,d}V} \int_{V_0} \Gamma u_f(x,y,z,t) \text{\,d}V,
\end{equation}
where the subscript $V$ of the angular brackets now indicates spatial averaging.
This is again followed by a central moving average.
Temporal averaging as in Eq.~\eqref{eq:temporal_averaging} finally yields $\langle u_f \rangle_{V,t}$, the vertical fluid profile consecutively averaged over space and time.
We note that we observed temporal fluctuations in the instantaneous flow profiles within the bulk of the sediment bed, i.e. where the fluid and particle velocities are very small.
Those fluctuations presumably originate from ongoing sorting effects inside the bed that appear over long time spans \citep{ferdowsi2017}. 
As such, longer simulation times would be desirable to increase the temporal averaging window and obtain a more robust statistical steady state. 
It was shown by \cite{vowinckel2020}, however, that unsteady effects are negligible when analyzing the rheological properties in the viscous regime. 

\begin{figure}
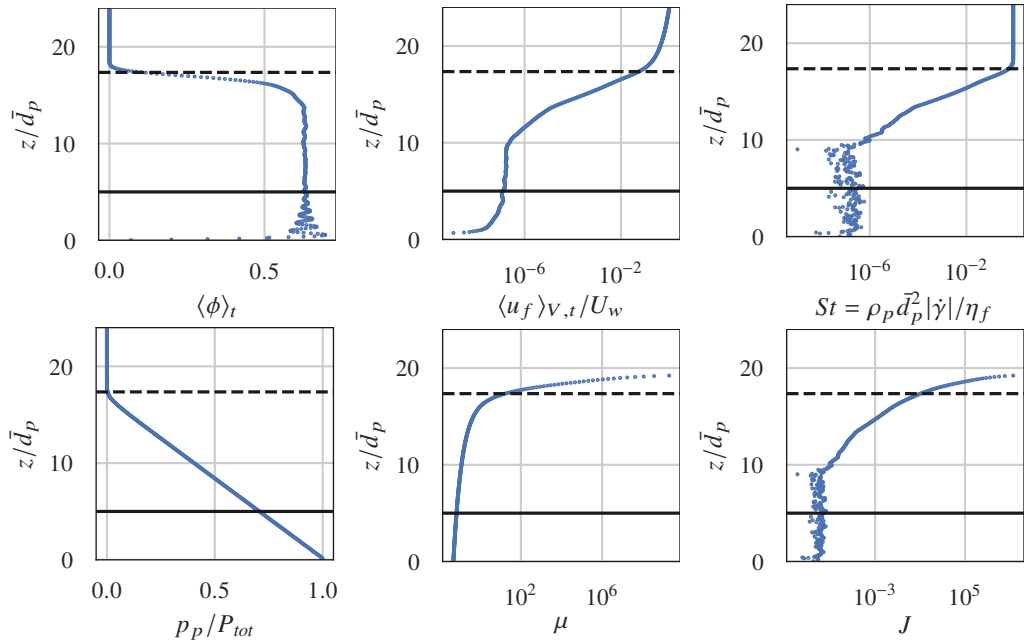

	\centering
	\begin{subfigure}[h]{0.33\textwidth}
		\centering
		\input{figures/profiles_timeaveraged_packingfraction_var01.pgf}
	\end{subfigure}~
	\begin{subfigure}[h]{0.33\textwidth}
		\centering
		\input{figures/profiles_timeaveraged_particlevelocity_var01.pgf}
	\end{subfigure}~
	\begin{subfigure}[h]{0.33\textwidth}
	\centering
	\input{figures/profiles_timeaveraged_shearrate_var01.pgf}
\end{subfigure}

\begin{subfigure}[h]{0.33\textwidth}
	\centering
	\input{figures/profiles_timeaveraged_particlePressure_var01.pgf}
\end{subfigure}~
\begin{subfigure}[h]{0.33\textwidth}
	\centering
	\input{figures/profiles_timeaveraged_frictionCoeff_var01.pgf}
\end{subfigure}~
\begin{subfigure}[h]{0.33\textwidth}
	\centering
	\input{figures/profiles_timeaveraged_viscnumber_var01.pgf}
\end{subfigure}
	\caption{Spatially and temporally averaged profiles of different quantities for the monodisperse case. The dashed horizontal line represents the bed height. The solid horizontal line is at $z=5\bar{d}_p$ and all profile data below is discarded in further analysis.
	The data of the vertical profiles for all simulation runs are provided as supplemental material.
	}
	\label{fig:evaluated_profiles_from_data}
\end{figure}

We obtain the local shear rate as the spatial derivative of $\langle u_f \rangle_{V,t}$. Owing to the spatial heterogeneity of our polydisperse sediment beds that may still be subject to ongoing sorting, we decided to use the absolute value of the local shear rate, i.e. $\lvert \dot{\gamma}\rvert$, as a robust measure to compute the rheological quantities \citep{madraki2017}. The actual shear stress $\tau$ is extracted from the bulk region of the flow, where it is constant due to the linear flow profile.
The normalized shear stress values of all cases are reported in table \ref{tab:setups_evaluation}, which are close to the target Shields number of $0.5$.
The granular pressure, on the other hand, is obtained from $\langle\phi\rangle_t$ via
\begin{equation}
p_p(z) = \left(\rho_p-\rho_f\right) g \int_{z}^{\infty} \langle\phi\rangle_t(z') \text{\,d}z'.
\end{equation}
This definition is in line with the one proposed by the two-phase model of \cite{aussillous2013} and successfully used in the analysis of \cite{vowinckel2020}. Note that we do not introduce an artificial confining pressure $P_0$ at the top wall as suggested by \cite{houssais2016}, because our simulation data yields full information of vertically resolved porosity profiles across the entire depth of the channel. 
These data allow for a straightforward computation of the vertical profiles of $\mu$ and $J$. 
The final profiles of the relevant quantities are exemplified in figure~\ref{fig:evaluated_profiles_from_data} by showing the results for the monodisperse case.
In this figure, the granular pressure is normalized by $P_\mathit{tot} = \left(\rho_p-\rho_f\right) g \int_{0}^{\infty} \langle\phi\rangle_t(z') \text{\,d}z'$, which is the total submerged weight of the sediment bed.
The complete data sets for all four simulation cases can be found in the supplementary data.
Looking at the particle volume fraction profile, a layering is visible near the bottom plane (figure~\ref{fig:evaluated_profiles_from_data}a), which is due to the ordered structure induced by the spheres mounted to the bottom plane.
Therefore, we discard the data from the lower parts of the bed, i.e. where $z < 5\bar{d}_p$, to exclude potential artefacts induced by the boundary condition of the bottom roughness.

We can directly obtain the maximum solid volume fraction $\phi_m$ from the particle volume fraction profile. To this end, we evaluate its average in the bulk region of the bed, i.e.
\begin{equation}
\phi_m = \frac{1}{5\bar{d}_p} \int_{5 \bar{d}_p}^{10 \bar{d}_p} \langle\phi\rangle_t(z) \text{\,d}z.
\end{equation}
Its value for the different setups is given in table \ref{tab:setups_evaluation}.  
As expected, $\phi_m$ increases with polydispersity since the voids between larger particles can be filled by smaller particles.
\RevThree{The maximum packing fractions are close to the values commonly reported in literature for random close sphere packings with log-normal size distributions~\citep{brouwers2014, farr2013}.}

For brevity, we will omit the indication of the averaging operator and use $\phi$ instead of $\langle\phi\rangle_t$ to denote the averaged particle volume fraction for the remainder of the work.

\section{Rheology of monodisperse sediment beds}
\label{sec:rheology_monodisperse}

\subsection{Rheological model for dense suspensions}

The rheology of monodisperse, neutrally buoyant, spherical particles in a viscous fluid has been assessed experimentally by shearing walls that impose a constant volume on the fluid particle mixture \citep[e.g][]{Krieger1959,morris1999,Stickel2005,guazzelli2011}.
This approach is commonly referred to as \textit{volume-imposed rheometry}.
The scenario has been extended to a \textit{pressure-imposed rheometry}, where a constant confining pressure is applied on the top wall that remains movable in the vertical direction. This measure allows to investigate the dilation/consolidation of a granular suspension under varying shear \citep[e.g.][]{boyer2011,Dagois2015,tapia2019}. 
As already laid out in the introduction, this scenario bears a straightforward analogy to the shearing of sediment beds.
Hence, the \textit{pressure-imposed rheometry} and the corresponding empirical correlations derived from the rheological experiments to predict the macroscopic friction and the particle volume fraction as functions of the viscous number $J=\eta_f \dot{\gamma}/p_p$ are the focus of this work.

Using their experimental apparatus, \cite{boyer2011} followed the argument of \cite{cassar2005} to show that the rheology of the fluid-particle mixture is governed by $J$.
Based on these considerations, \cite{boyer2011} proposed the following empirical correlations as a rheological model, which became known as the $\mu(J)$-rheology and reads in its most general form
\begin{align}
\mu(J) &= \underbrace{\mu_1 + \frac{\mu_2-\mu_1}{1+J_f/J}}_{\mu^f(J)}+ \underbrace{a_\mu J^{1/2} + b_\mu J}_{\mu^h(J)}, \label{eq:mu_j_rheology} \\ 
\phi(J) &= \frac{\phi_m}{1+(K_n J)^{1/2}}. \label{eq:phi_j_rheology}
\end{align}
The macroscopic friction coefficient, thus, has the two contributions $\mu^f$ and $\mu^h$ from frictional-contact-based and hydrodynamic stresses, respectively.
The expression of $\mu^f$ was originally proposed by \cite{jop2005} and \cite{cassar2005} while studying submarine granular flow down an inclined plane.
Notably, the parameter $J_f$ represents the value of $J$ for which $\mu^f = (\mu_2 + \mu_1)/2$, i.e. the average of $\mu_1$ and $\mu_2$.
This parameter can therefore be understood as the transition from a frictional dominated to a more suspended regime where binary particle collisions prevail and the role of hydrodynamic stress becomes increasingly important. The parameters $\mu_1$ and $\phi_m$ are particle properties that represent the minimum friction and maximum particle volume fraction, respectively, for $J\rightarrow 0$, i.e. the jamming point of the dense suspension when the granular flow ceases. According to \cite{cassar2005}, $\mu_2$ is the maximum value for the friction coefficient at higher shear rates, whereas, this value serves as the threshold that distinguishes the two contributions from particle contact and hydrodynamic interactions in the framework of \cite{boyer2011}.
The coefficients $a_\mu=1$, $b_\mu=5/2 \phi_m$ can be determined from the analytical solution for effective viscosities of dilute suspensions originating from \cite{Einstein1905}, and $K_n$ is a parameter that has been determined empirically by best fit to experimental data \citep{morris1999,boyer2011}.

For the sake of the arguments that follow, we decided to deviate from the commonly encountered notation of $J_f$, which \RevThree{has previously been denoted as $I_0$ \citep[e.g.][]{cassar2005,boyer2011,houssais2016} or $J_0$ \citep[e.g.][]{Guazzelli2018,vowinckel2020}}.

\subsection{Existing model parameterizations}

\begin{table}
    \centering
    \begin{tabular}{llllllll}
    \toprule
    work & range of $J$ & $\mu_1$ & $\mu_2$ & $J_f$ & $a_\mu$ & $b_\mu$ & $\phi_m$\\\midrule
    \cite{cassar2005} & $[10^{-5}, 10^{-1}]$(*) & 0.43 & 0.82 & 0.0027(*) & 0 & 0 & - \\
    \cite{boyer2011} & $[10^{-6}, 10^{-1}]$ & 0.32 & 0.7 & 0.005 & $\frac{5}{2} \phi_m$ & 1 & 0.585\\
    \cite{houssais2016} & $[3 \times 10^{-5}, 2]$ ($\dagger$) & 0.27 & 0.52 & 0.0012 & $\frac{5}{2} \phi_m$ & 1 & 0.589\\
	\cite{tapia2019} (SR)& $[3 \times 10^{-4}, 10^{-1}]$ & $0.37$ & $\mu_1$ & - & $5.45$ & 0 & $0.584$ \\
	\cite{tapia2019} (HR)& $[3 \times 10^{-4}, 10^{-1}]$ & $0.36$ & $\mu_1$ & - & $5.16$ & 0 & $0.565$ \\
    \bottomrule
    \end{tabular}
    \caption{Summary of previous work in the context of the $\mu(J)$-rheology, Eq.~\eqref{eq:mu_j_rheology}, together with the reported coefficients. (*): The values for $J$ and $J_f$ were adapt to match our definition of the viscous number. ($\dagger$): Range used for fitting.
    }
    \label{tab:existing_rheolgical_model_coeffs}
\end{table}

In the work of \cite{boyer2011}, viscous numbers in the range $J \in [10^{-6},10^{-1}]$ were investigated. 
Since $\lim_{J\rightarrow 0}\mu^f(J) = \mu_1$ and $\lim_{J\rightarrow 0}\phi(J) = \phi_m$,  the parameters $\mu_1=0.32$ and $\phi_m = 0.585$ were obtained within the lower limit of $J$.
Additionally, the parameters $\mu_2=0.7$ and $J_f=0.005$ were determined by fitting to the experimental data. 
The coefficient $K_n$ was evaluated as $K_n=1$ by \cite{boyer2011}, whereas \cite{morris1999} found a value of $K_n=0.75$ in their experiments on shear-induced particle migration. 

Recently, further experimental studies of an annular flume setup with monodisperse spheres were reported by \cite{houssais2016} and \cite{tapia2019}, which differ most notably in the range of measured $J$ values.
In \cite{houssais2016}, a sediment bed of monodisperse spheres was sheared by a laminar Couette flow to obtain values of $J \in [10^{-9},10]$, which extended the data range to significantly lower $J$. 
This study revealed a novel regime for $\mu$, labeled as the \textit{creep} regime and it is discussed in more detail in \S~\ref{sec:creep_extension}.
To provide a comparison with \eqref{eq:phi_j_rheology}, \cite{houssais2016} decided to exclude these low $J$-values from their analysis to obtain fitted coefficients for the region $J \in [3\times10^{-5},2]$ that show very good agreement with the results of \cite{boyer2011}. 

In contrast, \cite{tapia2019} investigated a region of $J \in [3\times10^{-4},10^{-1}]$ to address the effect of particle roughness on the rheology of dense suspensions.
For that reason, they used slightly roughened (SR) and highly roughened (HR) spheres in their experiments.
Instead of fitting the complete Boyer model \eqref{eq:mu_j_rheology}, these authors suggested a simplified scaling, which only contains the $\sqrt J$ term close to the jamming transition and used this approach to determine the friction factor at the jamming point by extrapolating their data.
This approach worked very well for the given range of $J$, but it also required a fitting of the coefficient $a_\mu$ that was, thus, found to be different from the Einstein formulation. 
Following the reasoning given in \cite{tapia2019}, they assumed a constant $\mu^f$ which implies $\mu_1 = \mu_2$.
This effectively removes the second term of $\mu^f$ from \eqref{eq:mu_j_rheology} and, thus, $J_f$ is not required for this analysis.

A summary of the values that have been reported in literature and discussed in the preceding paragraphs is given in table \ref{tab:existing_rheolgical_model_coeffs}. Note that the particles used in all of these experimental studies were monodisperse spheres.  

\subsection{Comparison to simulation results} \label{sec:mono_simulation}

\begin{figure}
	\centering
	\input{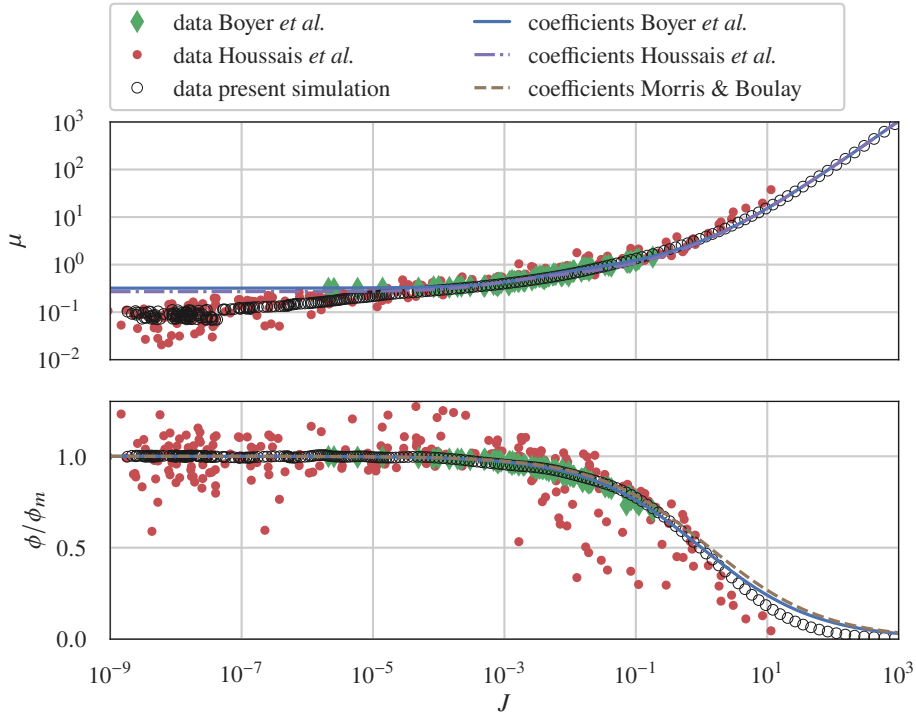}
	\caption{Rheological quantities (top: $\mu$, bottom: $\phi$) as function of viscous number $J$ for a monodisperse sediment bed. Data from the present monodisperse simulation is shown, together with experimental data from \cite{boyer2011} and \cite{houssais2016}. Additionally, curves of equations \eqref{eq:mu_j_rheology} and \eqref{eq:phi_j_rheology} are shown, parameterized as proposed by \cite{boyer2011,houssais2016,morris1999} (cf. table \ref{tab:setups_evaluation}).}
	\label{fig:result_mono_muPhi}
\end{figure}

In an effort to compare our simulation results against experimental data of {\em pressure-imposed rheometry}, we evaluate our data following the procedure described in \S\ref{sec:evaluation} to extract all rheological quantities as vertical profiles through the sediment bed (cf. figure \ref{fig:evaluated_profiles_from_data}).
Combining the data from these profiles, we are able to investigate $\mu$ and $\phi$ as a function of $J$ within the range $J \in [10^{-9},10^3]$.
This analysis is shown in figure \ref{fig:result_mono_muPhi} for the monodisperse case.
In the upper panel of this figure, the macroscopic friction factor $\mu$ is given as a function of the viscous number $J$. 
For comparison, we plot our data together with the experimentally obtained data from \cite{boyer2011} and \cite{houssais2016}, as well as the therein proposed parameterization of the $\mu(J)$ model \eqref{eq:mu_j_rheology} as summarized in table \ref{tab:existing_rheolgical_model_coeffs}.
The lower panel of the same figure shows our data for the particle volume fraction $\phi$ over $J$ normalized by $\phi_m$, and the predictions using \eqref{eq:phi_j_rheology} with the coefficient $K_n$ from \cite{boyer2011} and from \cite{morris1999}.

Comparing our simulation results of $\mu(J)$ to the existing experimental data shows a very good agreement, in particular with the data from \cite{houssais2016} over the complete range of $J$.
Consequently, the simulation data is well predicted by the parameterized models \eqref{eq:mu_j_rheology} for $J > 10^{-5}$. 
This range is in agreement with the values used in these experimental studies to calibrate the coefficients $\mu_1$, $\mu_2$, and $J_f$. For lower values of $J$, our data underestimates the two correlations, which confirms the {\em creep} regime reported by \cite{houssais2016} and visible in their data. 
In this regime, the plotted parameterizations of the model predict that $\mu$ levels off to a constant value, whereas the available data shows another significant shift towards a lower level of $\mu$.
 
The simulation results for $\phi(J)/\phi_m$ match well with the experimental data of \cite{boyer2011}, normalized by $\phi_m=0.585$, and \cite{houssais2016}, normalized by $\phi_m=0.589$. 
The latter shows some significant scatter, originating from the five distinct experiments varying the Shields numbers.
Excellent agreement between our data and the rheology model is observed for the range $J \in [10^{-9},1]$, which contains the range of viscous numbers used in \cite{boyer2011} to parameterize the model. 
For larger $J$, the simulation data exhibits smaller $\phi$ values than either of the models. 
In this range, we observe a more rapid decrease of $\phi$ from $\phi_m$ to 0. This region corresponds to the interface between the densely packed sediment bed and free flow region.  
The deviations reflect the difficulty to use the empirical correlation of \cite{boyer2011} in the extrapolated region of a more dilute regime \citep{vowinckel2020}. 
By comparing the two parameterizations, we see that the parameter $K_n$ in \eqref{eq:phi_j_rheology} controls the viscous number range of this transition region.
We note that the value of $\phi_m$, used for the normalization of our simulation data, is 0.631 and thus larger than the ones from other studies. As already noted \S\ref{sec:evaluation}, our value of $\phi_m$ is close to the one reported for a random sphere packing which can be expected since it is obtained from the bulk region of the sediment bed, i.e., the region of vanishingly low shear rates and, consequently, small viscous numbers.
This is in contrast to other studies \citep{boyer2011,vowinckel2020}, where stronger shearing was applied that led to a notable dilation of the suspension and, thus, a decrease in $\phi_m$.
\RevTwo{Furthermore, \cite{singh2018} observed a strong influence of the inter-particle friction coefficient $\mu_p$ on $\phi_m$ for sheared systems and found values of $\phi_m$ that are similar to ours for a friction coefficient of $\mu_p=0.15$.}
To focus on the general behavior of the $\phi(J)$ relation rather than the limiting value, which is therefore different in our simulation but also in existing studies, we always present and analyze the normalized $\phi$ values in this work.
This also effectively removes the dependence on $\phi_m$ from the the $\phi(J)$ model \eqref{eq:phi_j_rheology}.

In summary, our data of the monodisperse case agrees well with existing experimental data and previously derived parameterizations of the rheology model.
This overall confirms the validity of our simulation approach for densely packed sediment beds in shear flow and enables further predictive simulations.
These studies will feature polydisperse setups for direct comparison with the monodisperse models.
Furthermore, we observe a systematic shift in $\mu$ towards lower values for $J<10^{-5}$, also present in the experimental data of \cite{houssais2016}.
This range, however, was not addressed by \cite{boyer2011} nor \cite{houssais2016} and is thus not contained in the existing rheological model.
In the following section, we will evaluate and enhance the parameterization of the empirical coefficients in \eqref{eq:mu_j_rheology} for the effects of polydispersity by focusing on the collisional and hydrodynamic regime for $J \in [10^{-5},10^2]$.  We then proceed in \S\ref{sec:creep_extension} to study the {\em creep} regime in more detail and propose an extended model that is able to capture the observed behavior.

\section{Rheological model for polydisperse sediment beds}
\label{sec:polydisperse_extension}

\subsection{Simulation results}

\begin{figure}
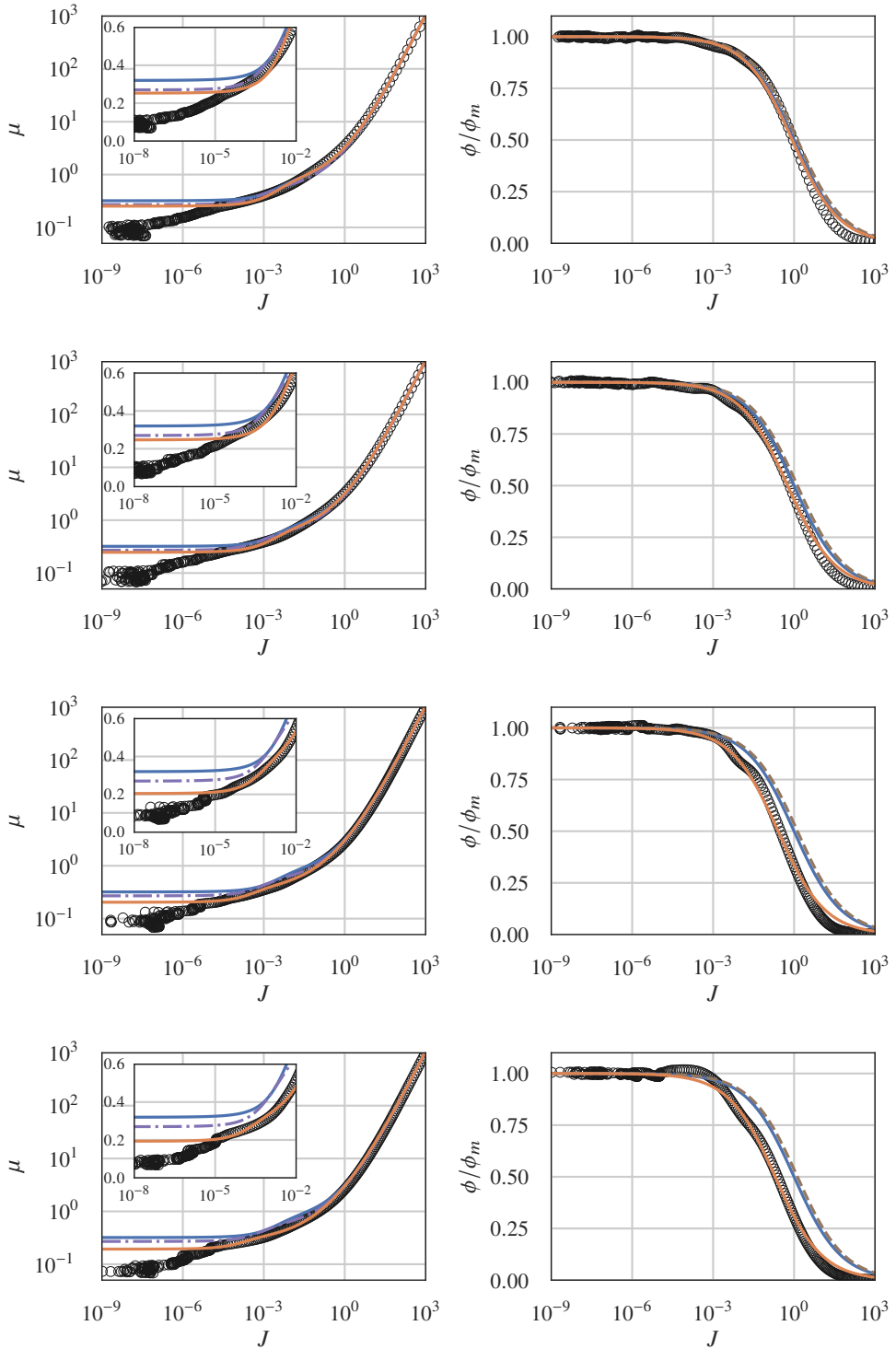

	\centering
	\begin{subfigure}{\textwidth}
		\centering
		\input{figures/eval_over_viscNumber_withFit_var01.pgf}
	\end{subfigure}
	\begin{subfigure}{\textwidth}
		\centering
		\input{figures/eval_over_viscNumber_withFit_var10.pgf}
	\end{subfigure}
	\begin{subfigure}{\textwidth}
		\centering
		\input{figures/eval_over_viscNumber_withFit_var50.pgf}
	\end{subfigure}
	\begin{subfigure}{\textwidth}
		\centering
		\input{figures/eval_over_viscNumber_withFit_var100.pgf}
	\end{subfigure}
	\caption{Rheological quantities (left: $\mu$, right: $\phi$) as function of viscous number $J$ for the four different setups (from top to bottom: \mono, \polya, \polyb, \polyc). Color and style as in figure \ref{fig:result_mono_muPhi}. Additionally, best fits as explained in \S\ref{sec:poly:fit} are given as orange curves. The insets in the left column magnify the region of small viscous numbers, using a linear axis for $\mu$.}
	\label{fig:result_muPhi_viscousnumber}
\end{figure}

We now apply the same analysis as for the monodisperse case in \S\ref{sec:mono_simulation} for the additional three setups of polydisperse sediment beds summarized in table \ref{tab:bed_properties} that reflect different degrees of polydispersity as indicated by the variance of the grain size distribution.
This analysis again yields $\mu$ and $\phi$ as a function of $J$ and is shown in figure \ref{fig:result_muPhi_viscousnumber}.

Similar to figure \ref{fig:result_mono_muPhi}, the left column shows the macroscopic friction factor $\mu$ from our data together with the model parameterizations from \cite{boyer2011} and \cite{houssais2016}.
For increasing polydispersity, we observe a decrease of $\mu$ within the range $J \in [10^{-5},10^0]$.
Note that $\mu$ and $J$ are plotted on logarithmic scales, i.e. even small deviations that become visible in this range are large in actual values, as can be seen in the respective insets.
All cases reproduce the {\em creep} regime for $J < 10^{-6}$, as already observed for the monodisperse case. 
This effect becomes slightly more pronounced with increasing polydispersity. 

The right column of figure \ref{fig:result_muPhi_viscousnumber} shows our data for the particle volume fraction $\phi$ over $J$ normalized by $\phi_m$, and model parameterizations from \cite{boyer2011} and \cite{morris1999}.
There, the drop from $\phi_m$ to 0 occurs at lower values of $J$ when the polydispersity is increased, which results in a shift by up to one order of magnitude in $J$ for \polyc{} compared to \mono. 
An interesting feature emerges for values of $\phi$ around $J\approx 10^{-4}$ that can be seen most prominently for the \polyc{} case where values larger than $\phi_m$ are observable.
We found this to be a result of vertical sorting of the polydisperse sediment, where finer sediments from the topmost sediment layer translate to and accumulate in a lower layer, thereby increasing the particle volume fraction in this region.

Summarizing, increasing the polydispersity of the sediment bed while keeping all other physical parameters constant has a distinct effect on $\mu$ and $\phi$ as a function of $J$.
As a result, the agreement between the simulation data and the existing model parameterizations by \cite{boyer2011}, \cite{houssais2016}, and \cite{morris1999} deteriorates with increasing polydispersity.
In the following, we will enhance the parameterization of the rheological model in \eqref{eq:mu_j_rheology} and \eqref{eq:phi_j_rheology} for the effects of polydispersity by focusing on the frictional and hydrodynamic regime for $J \in [10^{-5},10^2]$.
For now, we exclude the {\em creep} regime for the remainder of this section to provide a consistent comparison with the analyses of \cite{boyer2011} and \cite{houssais2016}. However, we will study this regime in more detail in the subsequent section \S\ref{sec:creep_extension}.

\subsection{Effect of polydispersity on model parameterization} \label{sec:poly:fit}

\begin{table}
	\centering
	\begin{tabular}{llcccc}
	\toprule
		&&  \multicolumn{3}{c}{Eq.~\eqref{eq:mu_j_rheology}} & Eq.~\eqref{eq:phi_j_rheology} \\\cmidrule(lr){3-5} \cmidrule(lr){6-6}
		 & & $\mu_1$& $\mu_2$ & $J_f$ & $K_n$ \\\midrule
		present fits:&&&&&\\
		&\mono & 0.253 & 0.704 & 0.0059 & 1.165 \\
		&\polya & 0.247 & 0.577 & 0.0041 & 1.743 \\
		&\polyb & 0.204 & 0.367 & 0.0006 & 3.896 \\
		&\polyc & 0.193 & 0.301 & 0.0002 & 4.982 \\
		others: &&&&&\\
		&\cite{morris1999}  & - & - & - & 0.75 \\
		&\cite{boyer2011}  & 0.32 & 0.70 & 0.0050 & 1 \\
		&\cite{houssais2016}  & 0.27 & 0.52 & 0.0012 & - \\
		\bottomrule
	\end{tabular}
	\caption{Coefficients applied for the equations of the $\mu(J)$ and $\phi(J)$ rheology for the curves shown in figure \ref{fig:result_muPhi_viscousnumber}, with $\phi_m$ from table \ref{tab:setups_evaluation}. The fits are obtained using data of $J\in [10^{-5},10^2]$.}
	\label{tab:fitted_model_coefficients}
\end{table}

In order to improve the parameterization of equations \eqref{eq:mu_j_rheology} and \eqref{eq:phi_j_rheology}, we evaluate the parameters $\mu_1, \mu_2, J_f$, and $K_n$ determined from fits of our simulation results to reveal trends as a function of increasing polydispersity.
To this end, we apply a fit of \eqref{eq:mu_j_rheology} and \eqref{eq:phi_j_rheology} to our data.
We follow the reasoning of \cite{boyer2011} and determine $\mu_1$, $\mu_2$, and $J_f$ as free parameters, while keeping $a_\mu=5/2\phi_m$ and $b_\mu=1$ to recover the Einstein relation for the effective viscosity of dilute suspensions. 
Similar to \cite{houssais2016}, we apply the fit over the range $J \in [10^{-5},10^2]$ and exclude the values for lower $J$ to focus on the regimes dominated by frictional and hydrodynamic stresses.
Owing to the large value range over several orders of magnitude, we fit $\ln(\mu)$ to $J$ instead of $\mu$ directly. 
The resulting coefficients are reported in table \ref{tab:fitted_model_coefficients}, and the corresponding plots are additionally presented in figure \ref{fig:result_muPhi_viscousnumber}.
We explicitly note that $\phi_m$ is extracted from our simulation results as a quantity of the individual sediment bed and is not fitted here.

Comparing the case \mono{} to \cite{boyer2011}, our values for $\mu_2$ and $J_f$ are almost identical, and $K_n$ also agrees very well, but we found a value for $\mu_1$ that is closer to the results of \cite{houssais2016}. 
This could be attributed to the material parameters that enter our particle contact algorithm described in \S\ref{sec:DEM}, such as the restitution coefficient and friction coefficient, which are parameters that are not reported by neither one of these experimental studies. 

For increasing polydispersity, the friction coefficients $\mu_1$ and $\mu_2$ decrease, while $K_n$ increases.
Additionally, $J_f$ changes in the four cases as well, although the values remain on a very low level for all cases. A significant shift was detected from $J_f=0.0042$ to $J_f=0.0006$  for the cases \polya{} and \polyb, respectively, whereas $J_f$ remains on this lower level for \polyc{}, 
Owing to the large range of $J$, it is challenging to precisely determine the exact value of $J_f$ via curve fitting. 

In the case of $\phi$, the fitted curves reproduce the position and extent of the drop from $\phi_m$ to 0 particle volume fraction very well.
This is achieved by increasing $K_n$ for larger polydispersities, resulting in significantly larger values than given by \cite{morris1999} and \cite{boyer2011}.
Slight deviations of the simulation data from the fitted correlations can still be seen for $J \approx 10^2$ where the curves predict values larger than present in the data.

Generally, the fitted curves plotted in figure \ref{fig:result_muPhi_viscousnumber} show a very good agreement with the simulation results for the here considered range of viscous numbers.
This confirms our assumption that an adequate parameterization of the existing models for $\mu(J)$ and $\phi(J)$ allows for an extension that takes polydispersity into account.
In a next step, we attempt to formalize the observed trends in the obtained coefficients as functions of polydispersity.

\subsection{Model parameterization as a function of polydispersity} \label{sec:poly:model}

\begin{figure}
	\centering
	\input{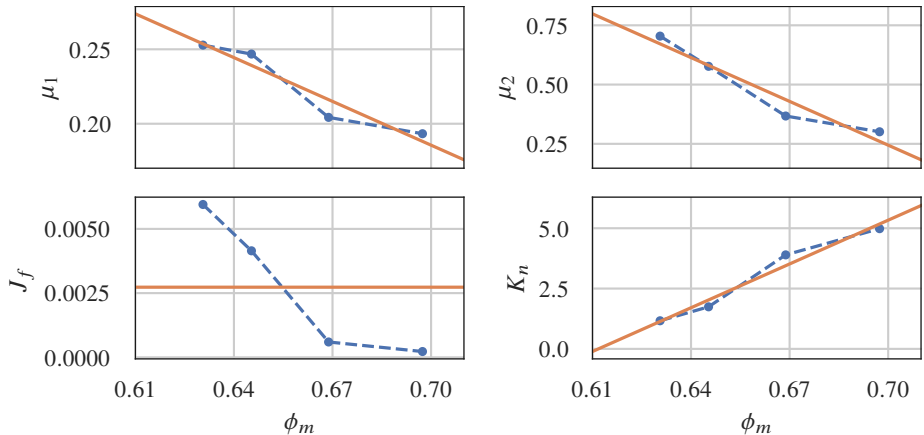}
	\caption{Fitted coefficients (blue) from table \ref{tab:fitted_model_coefficients} as function of parameter $\phi_m$, which is used to describe polydispersity. Additionally, the correlations \eqref{eq:model_mu1}-\eqref{eq:model_Kn} are included as orange lines.}
	\label{fig:model_muphi_coefficients}
\end{figure}

From the fits to the four different simulated cases, we find that the coefficients entering \eqref{eq:mu_j_rheology} and \eqref{eq:phi_j_rheology} depend on the polydispersity of the sediment bed.
The parameters $\mu_1$ and $\mu_2$ decrease when the polydispersity is increased, whereas $K_n$ increases.
Even though $J_f$ seemingly decreases with increasing polydispersity, we refrain from interpreting these values as an actual trend due to the aforementioned difficulties in its determination. 
Based on these findings, we aim to extend the existing rheological model to incorporate polydispersity in a general way and without individual calibration or fitting.
As such, it becomes readily applicable in macroscopic simulations and can significantly improve the predictions of the rheology of polydisperse sediment beds. 

To this end, we have to select a parameter that characterizes polydispersity in a concise way.
A set of possible parameters can be found in table \ref{tab:bed_properties}, \RevOne{namely the variance of the underlying log-normal distribution as well as the the diameter ratio $d_{p,\mathit{max}}/d_{p,\mathit{min}}$.} It is also reported in table \ref{tab:setups_evaluation} that these parameters directly influence the maximum particle volume fraction $\phi_m$ that indicates the jamming condition. 
Here, we choose $\phi_m$ to be the characteristic parameter as it is already present in the existing rheological framework as a key parameter.
\Rev{This choice of the governing parameter is in line with recent work by \cite{pednekar2018} and the quantity can be obtained in a robust manner from either the vertical profile of the particle volume fraction or from $\phi(J)$ as $J\rightarrow 0$.}
\RevOne{For an a priori determination of $\phi_m$, a reasonable estimation can be obtained by assuming a perfect log-normal distribution and making use of available packing fraction predictors \citep[e.g.][]{brouwers2014,farr2013}.}
\RevOne{Previous studies on dry granular flows have suggested to account for polydispersity by using the weighted arithmetic mean
of the particle diameter in the definition of the inertial number \citep{tripathi2011}. Since this geometric quantity does not appear in the definitions of $\mu(J)$-rheology framework, we identified $\phi_m$ as the more suitable measure to account for polydispersity of dense suspensions in a quantitative manner.}
In figure \ref{fig:model_muphi_coefficients}, the fitted coefficients are plotted as a function of $\phi_m$. 

In a next step, a functional expression for each parameter is determined which describes the dependence on $\phi_m$.
For the three parameters with a clear trend, we assume a linear dependence on $\phi_m$. 
This is the strongest assumption we can justify based on the number of data points available.
For $J_f$, we refrain from further assumptions and use the average of the fitted values, while also reporting its standard deviation.
\RevOne{
A sensitivity study revealed that the dependence on the exact value of $J_f$ is only weak, so that solely its order of magnitude, which is captured well by the average, has a significant effect.
This justifies the model simplification and keeps the number of coefficients to a minimum. 
}
Applying a linear regression, the resulting correlations for each parameter are given as
\begin{align}
\mu_1 &= -0.980 \, \phi_m +0.872,\label{eq:model_mu1}\\
\mu_2 &= -6.157 \, \phi_m +4.554,\label{eq:model_mu2}\\
J_f    &= 0.0027 \pm 0.0024,\label{eq:model_Jf}\\
K_n    &= 60.490 \, \phi_m -37.008.\label{eq:model_Kn}
\end{align}
The above relations are plotted as orange solid lines in figure \ref{fig:model_muphi_coefficients} as well, exhibiting a reasonable agreement to the values determined by individual fits.

For a quantitative comparison, we assess the predictive power of the rheology model \eqref{eq:mu_j_rheology} and \eqref{eq:phi_j_rheology} to reproduce our observed simulation results using the parameters $\mu_1$, $\mu_2$, $J_f$ and $K_n$ proposed by \cite{boyer2011}, \cite{houssais2016} and \cite{morris1999}, as well as the ones found by the individual fits performed in \S \ref{sec:poly:fit} and compare it against the prediction using the parameterization given by the calibrated expressions \eqref{eq:model_mu1}-\eqref{eq:model_Kn}. 
To this end, we compute the $R^2$ value as measure to quantify the agreement between observations $o$ and a prediction model $m$ as
\begin{equation}
    R^2 = 1- \frac{\sum_i (o_i - m_i)^2}{\sum_i (o_i - \bar{o})^2},
\end{equation}
where $\bar{o}$ is the average value of all observations. 
The maximum $R^2=1$, thus, indicates perfect agreement between the model prediction and the observations, whereas smaller values mean lower agreement.

The $R^2$ values are reported in table \ref{tab:fitted_model_coefficients_r2}, where again we use the logarithmized data to compute $R^2$ for $\mu$ due to its large value range.
Note that we evaluated the $R^2$ for the range of $J \in [10^{-5},10^2]$, which corresponds to the value range used for fitting and excludes the creep regime.
For $\mu(J)$, the parameterizations from \cite{boyer2011} and \cite{houssais2016} offer a fairly good predictive quality for the monodisperse case and then deviate for increasing polydispersity, which is in line with our previous observations.
This is improved when applying the fitted coefficients which produces an almost perfect agreement in all four cases.
Our expressions for $\mu_1$, $\mu_2$, and $J_f$, \eqref{eq:model_mu1}-\eqref{eq:model_Jf}, yield a performance very similar to the fitted parameters.
In particular, this shows that the results are rather insensitive to the actual choice of $J_f$ as the values differ by one order of magnitude in the case of \polyc, which can be seen as an additional justification for assuming a constant $J_f$.
The same findings regarding the predictive quality can be reported for the particle volume fraction $\phi$. The individual fits and the correlation for $K_n$, \eqref{eq:model_Kn}, yield very good agreement for all the cases, whereas the parameterization by \cite{boyer2011} and \cite{morris1999}, i.e. $K_n= 1$ and $K_n= 0.75$, respectively, are not as accurate.

\begin{table}
	\centering
	\begin{tabular}{lcccccccc}
	\toprule
		& \multicolumn{4}{c}{$R^2\big(\ln(\mu(J))\big)$} & \multicolumn{4}{c}{$R^2\big(\phi(J)/\phi_m\big)$}  \\ \cmidrule(lr){2-5} \cmidrule(lr){6-9}
		& \multicolumn{2}{c}{others} & \multicolumn{2}{c}{present} & \multicolumn{2}{c}{others} & \multicolumn{2}{c}{present} \\\cmidrule(lr){2-3}\cmidrule(lr){4-5}\cmidrule(lr){6-7}\cmidrule(lr){8-9}
		case & Boyer & Houssais & fit & correlation & Boyer & Morris & fit & correlation \\\midrule
		\mono  &  0.990 & 0.994 & 0.998 & \textbf{0.997} & 0.994 & 0.985 & 0.995 & \textbf{0.995} \\
		
		\polya &  0.990 & 0.995 & 0.999 & \textbf{0.998} & 0.981 & 0.961 & 0.996 & \textbf{0.995} \\
		
		\polyb &  0.984 & 0.992 & 0.997 & \textbf{0.997} & 0.923 & 0.889 & 0.993 & \textbf{0.993} \\
		
		\polyc &  0.977 & 0.988 & 0.996 & \textbf{0.993} & 0.892 & 0.852 & 0.995 & \textbf{0.994} \\
		
		\bottomrule
	\end{tabular}
	\caption{$R^2$ values for different parameterizations of the rheology model, \eqref{eq:mu_j_rheology} and \eqref{eq:phi_j_rheology}, evaluated with respect to the simulated data for $J \in [10^{-5},10^2]$, thus excluding the creep regime. Present contributions consist of individual fits for each case with coefficients from table \ref{tab:fitted_model_coefficients}, and the novel correlations \eqref{eq:model_mu1}-\eqref{eq:model_Kn} taking into account polydispersity.}
	\label{tab:fitted_model_coefficients_r2}
\end{table}

From these results, we conclude that our approach of including the effect of polydispersity via a functional dependence of the coefficients on $\phi_m$ successfully improves the macroscopic rheology models.
Since the maximum particle volume fraction already appears in the original model, this strategy can readily be integrated and applied in macroscopic modeling approaches.

For $\mu(J)$, however, the region of small $J$, and accordingly small $\mu$, values can not be captured via the present formulation of \eqref{eq:mu_j_rheology}. 
As such, the applicability would be limited to cases with $J>10^{-5}$. 
To solve this issue, the model for $\mu(J)$ has to be extended to explicitly account for the {\em creep} regime as will be detailed in the next section. 
The model for $\phi(J)$, on the other hand, correctly predicts a constant value of $\phi_m$ for these small viscous numbers and is thus already applicable to this regime without further modifications.  

\section{Rheological model for creep regime}
\label{sec:creep_extension}

\subsection{Evaluation of the creep regime} 

The {\em creep} regime is characterized as a slow deformation of granular material under very low shear rates. In terms of the $\mu(J)$-rheology, this becomes evident by a macroscopic friction factor that does not level off to a constant value in the frictional regime, but decreases to even smaller values for lower and lower viscous numbers. Assessing this regime is challenging, because it requires very low viscous numbers. In fact, to the knowledge of the authors, the only experimental campaign that was able to investigate the rheology of the {\em creep} regime for granular flows immersed in a viscous shearing fluid is the study of \cite{houssais2016}, who reported values down to $J = 10^{-9}$. 
However, their results are subject to a substantial amount of scatter in this range due to the general difficulty of measuring such small $J$ and $\mu$ in an experimental apparatus that cannot be fully shielded from external disturbances and may touch the sensor accuracy of the measurement instruments.
Additionally, this study was carried out in an annular flume that introduces some artifacts due to the curved side walls. In our simulations of a straight horizontal domain with no side walls being present and the ability to control and evaluate the setup very accurately, these experimental imperfections are not an issue.
Despite the differences in the experimental setup of \cite{houssais2016} and our numerical simulation, we confirm the observation of the {\em creep} regime in our simulation data, as seen in figure~\ref{fig:result_muPhi_viscousnumber}, albeit with less scatter. This is true not only for the monodisperse case, that yields very good agreement with the experimental data of \cite{houssais2016} across the entire range of $J$ (figures \ref{fig:result_mu_viscousnumber_extended}a and \ref{fig:result_mono_muPhi}), but also for all other cases considered (figures \ref{fig:result_mu_viscousnumber_extended}b-d).

\begin{figure}
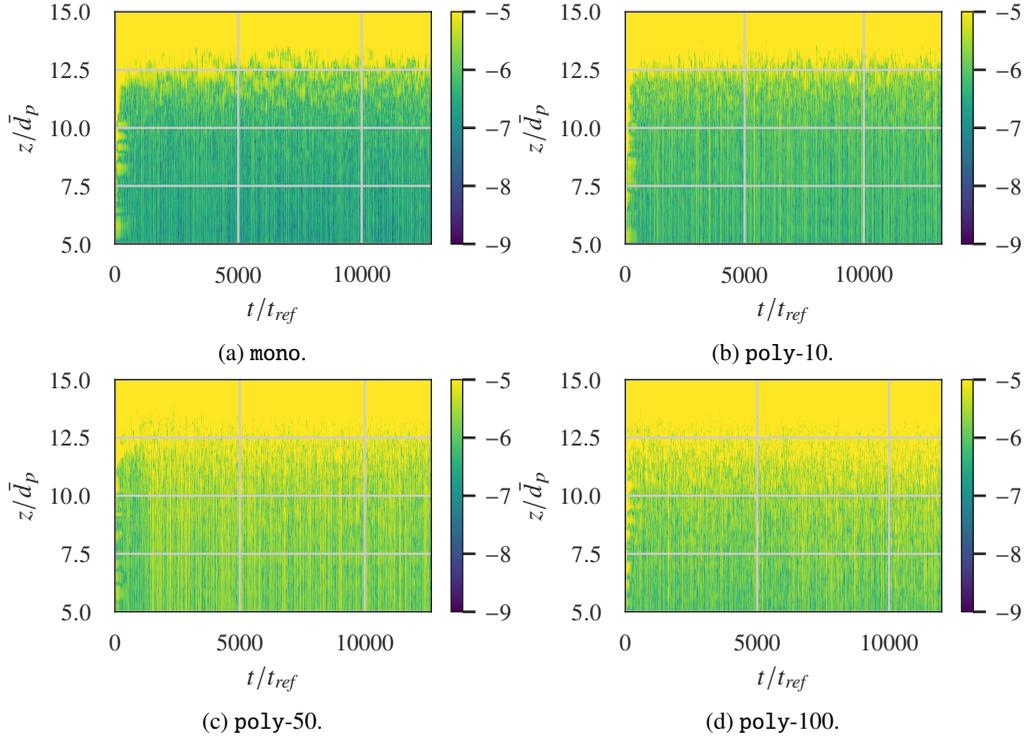

	\centering
	\begin{subfigure}{0.5\textwidth}
		\centering
		\input{figures/J_over_time_var01.pgf}
		\caption{\mono{}.}
	\end{subfigure}%
	\begin{subfigure}{0.5\textwidth}
		\centering
		\input{figures/J_over_time_var10.pgf}
		\caption{\polya{}.}
	\end{subfigure}
	\begin{subfigure}{0.5\textwidth}
		\centering
		\input{figures/J_over_time_var50.pgf}
		\caption{\polyb{}.}
	\end{subfigure}%
	\begin{subfigure}{0.5\textwidth}
		\centering
		\input{figures/J_over_time_var100.pgf}
		\caption{\polyc{}.}
	\end{subfigure}
	\caption{\RevTwo{Temporal evolution of vertical $J$-profiles. Due to the range of values, we plot $\log_{10}(J)$ to indicate the order of magnitude and choose the color scale to focus on very low viscous numbers.}}
	\label{fig:creep_temporal_J}
\end{figure}

\RevTwo{
\cite{houssais2016} perceived creep as localized, intermittent particle motion for which a description with temporally averaged quantities like $J$ and $\mu$ might be less appropriate.
To gain more insight into the dynamics of the creep regime and its mechanisms, we turn to the instantaneous but still spatially-averaged profiles of $J$.
These are visualized over time in figure~\ref{fig:creep_temporal_J} for all four simulated cases.
Note that the displayed vertical region is restricted to $z\in [5,15]\bar{d}_p$ to better focus on the creep regime.
Furthermore, we plot the viscous number in terms of $\log_{10}J$ due to its large value range. 
In all cases, we observe a short start-up phase which is followed by a statistically stationary state with temporal as well as vertical fluctuations.
These steady fluctuations agree qualitatively well with the ones reported for hard particles by \cite{bouzid2015}, who carried out two-dimensional simulations of sheared dry systems in the quasi-static limit.
This observation is in line with the particle properties used in our study, where the restitution coefficient and the particle friction were chosen to reflect silica grains.
Similar to the results by \cite{bouzid2015}, no burst-like behavior can be observed in figure \ref{fig:creep_temporal_J}.
On the contrary, \cite{bouzid2015} observed such intermittent motion only for soft particles with restitution coefficients as low as 0.1, which could then be better described by a non-local rheology~\citep[e.g.][]{kamrin2012}.
}

\begin{figure}
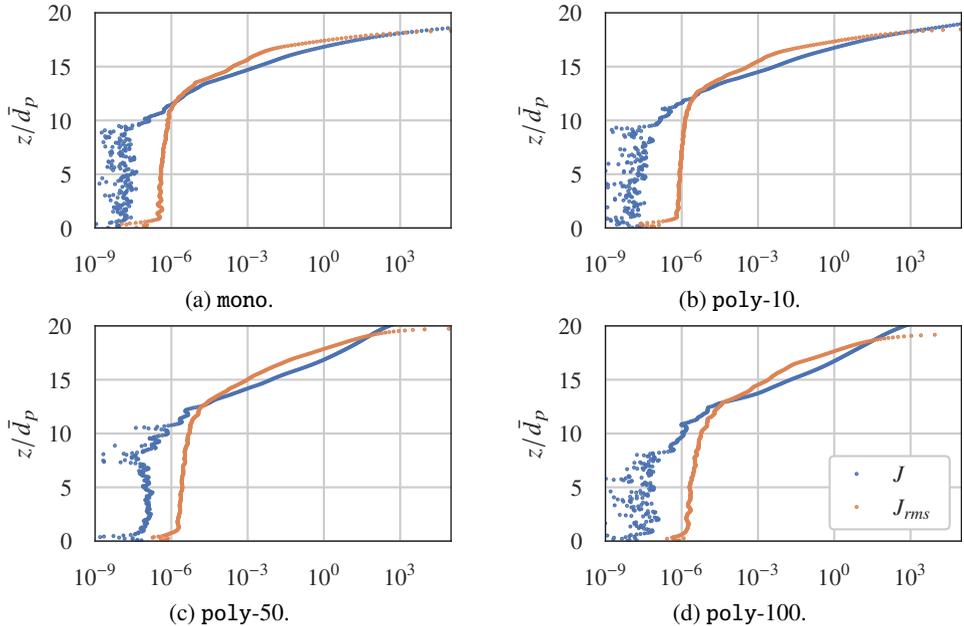

	\centering
	\begin{subfigure}{0.5\textwidth}
		\centering
		\input{figures/Jrms_over_height_var01.pgf}
		\caption{\mono{}.}
	\end{subfigure}%
	\begin{subfigure}{0.5\textwidth}
		\centering
		\input{figures/Jrms_over_height_var10.pgf}
		\caption{\polya{}.}
	\end{subfigure}
	\begin{subfigure}{0.5\textwidth}
		\centering
		\input{figures/Jrms_over_height_var50.pgf}
		\caption{\polyb{}.}
	\end{subfigure}%
	\begin{subfigure}{0.5\textwidth}
		\centering
		\input{figures/Jrms_over_height_var100.pgf}
		\caption{\polyc{}.}
	\end{subfigure}
	\caption{\RevTwo{Vertical profiles of the time-averaged $J$ and the root-mean-square (rms) value of its fluctuations, evaluated over the same time span $[t_0,t_1]$ as the temporal averaging given in table~\ref{tab:bed_properties}.}}
	\label{fig:creep_J_rms}
\end{figure}

\RevTwo{
Recently, \cite{gillissen2020} showed that temporal fluctuations of $J$, rather than its average, characterize the creep regime for inhomogeneous flow conditions.
These fluctuations are seen as the reason why the $\mu(J)$-rheology by \cite{boyer2011}, derived for homogeneous conditions, fails to capture the creep regime. 
Even though our considered setup is a homogeneous shear flow, we also observe significant fluctuations in this region of the bed.
We, therefore, follow the same argument and evaluate the vertical root-mean-square profile $J_\mathit{rms}$.
It is based on the deviations of the vertical instantaneous $J$ profiles from the temporally averaged one, evaluated over the same time span as the temporal average (excluding the initial start-up phase, cf. table~\ref{tab:bed_properties}).

This analysis of the vertical profiles of $J$ and $J_\mathit{rms}$ is shown in figure~\ref{fig:creep_J_rms} for the four simulated cases.
We observe that for viscous numbers above $10^{-6}$ (\mono) to around $10^{-5}$ (\polyc), the fluctuations are smaller than the average $J$. 
This is in agreement with results reported by \cite{gillissen2020} for homogeneous shear, and thus similar flow conditions.
Furthermore, this range corresponds to the viscous numbers, for which the existing $\mu(J)$-rheology was found to agree well with our simulation data, see \S \ref{sec:polydisperse_extension}.
Turning towards the {\em creep} regime, corresponding to the lower layers of the bed, the fluctuations exceed the averaged value by around two orders of magnitude. 
This was not observed by \cite{gillissen2020} for the case of homogeneous shear flow, as they could not access such small viscous numbers, so that the focus of this study was on inhomogeneous, and rather distinct, flow conditions of a Kolmogorov flow.
Interestingly, our evaluation also shows that the fluctuations surpass the average at larger viscous numbers of around $10$ as well. 
This coincides with the bed load transport layer at the fluid-sediment interface and is the region where the particles move along the bed's surface in an intermittent fashion, as they temporarily get trapped between particles and then proceed to slide or roll over them.

While the magnitude of these fluctuations thus might provide additional insight into the mechanisms of the creep regime, we note that the development of such rheological models is still an active field of research~\citep{gillissen2020}.
In particular, information about these fluctuations is usually not available in two-phase models and would require additional closure relations to be applicable there. 
Instead, we focus on the steady-state rheology and aim to include the creep regime as an extension to the existing $\mu(J)$-rheology in the next sections.
}

\subsection{Extension of model to creep}

Since the data by \cite{boyer2011} did not access such low viscous numbers, the description of this regime is, hence, lacking in the $\mu(J)$-rheology. To this end, we follow the reasoning of \cite{cassar2005} and \cite{jop2005}, and define a {\em creep} regime in addition to the frictional and hydrodynamic regime. Similarly to the frictional regime, this brings a lower and an upper limit of macroscopic friction, so that there remains a smooth transition in between the different regimes. This yields the following extension of equation \eqref{eq:mu_j_rheology} to adequately capture the {\em creep} regime in the rheological framework
\begin{equation}
    \mu(J) = \underbrace{\mu_0 + \frac{\mu_1-\mu_0}{1+J_c/J}}_{\mu^c} + \underbrace{\frac{\mu_2-\mu_1}{1+J_f/J}}_{\mu^f}+ \underbrace{\frac{5}{2} \phi_m J^{1/2} + J.}_{\mu^h} \label{eq:mu_j_extended_rheology}
\end{equation}
In comparison with the original model of \cite{boyer2011}, Eq.~\eqref{eq:mu_j_rheology}, we have shifted the lower limit of the macroscopic friction from $\mu_1$ to $\mu_0$, whereas $\mu_1$ becomes the upper limit of the creeping regime that centers around the viscous number of the creep regime, i.e., $J_c$. 
The proposed extension \eqref{eq:mu_j_extended_rheology} recovers the original formulation \eqref{eq:mu_j_rheology} by choosing $J_c=0$ or $\mu_0 = \mu_1$.
\RevTwo{We explicitly note that we here aim to model the rheological behavior for very small, but non-zero viscous numbers, i.e., $J\rightarrow 0$.
This quasi-static, but still dynamic, regime might thus be different from the static case at $J=0$ \citep{perrin2019}.}

\subsection{Testing the extended model for the creep regime}

\begin{figure}
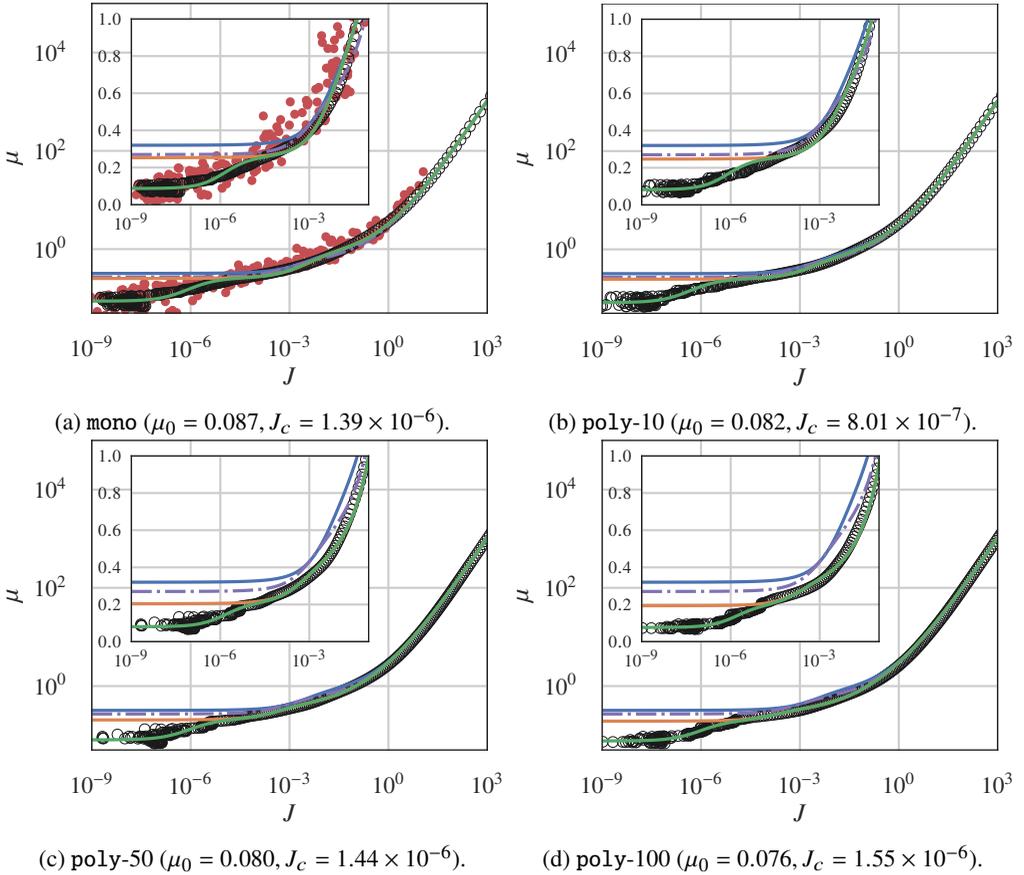

	\centering
	\begin{subfigure}{0.5\textwidth}
		\centering
		\input{figures/eval_over_viscNumber_withExtended_var01.pgf}
		\caption{\mono{} ($\mu_0 = 0.087, J_c = 1.39\times 10^{-6}$).}
	\end{subfigure}%
	\begin{subfigure}{0.5\textwidth}
		\centering
		\input{figures/eval_over_viscNumber_withExtended_var10.pgf}
		\caption{\polya{} ($\mu_0 = 0.082, J_c = 8.01\times 10^{-7}$).}
	\end{subfigure}
	\begin{subfigure}{0.5\textwidth}
		\centering
		\input{figures/eval_over_viscNumber_withExtended_var50.pgf}
		\caption{\polyb{} ($\mu_0 = 0.080, J_c = 1.44\times 10^{-6}$).}
	\end{subfigure}%
	\begin{subfigure}{0.5\textwidth}
		\centering
		\input{figures/eval_over_viscNumber_withExtended_var100.pgf}
		\caption{\polyc{} ($\mu_0 = 0.076, J_c = 1.55\times 10^{-6}$).}
	\end{subfigure}
	\caption{Macroscopic friction factor $\mu$ as function of viscous number $J$. 
	Legend as in figure \ref{fig:result_muPhi_viscousnumber}. In addition, the fit of the extended model, \eqref{eq:mu_j_extended_rheology}, is shown in green. The insets show a magnified view for low values of $J$ using a linear $y$-axis. 
	}
	\label{fig:result_mu_viscousnumber_extended}
\end{figure}

Similar to \S\ref{sec:poly:fit}, we apply curve fitting to find appropriate values for the newly introduced coefficients $\mu_0 \in [0,1]$ and $J_c$ for all simulations conducted.
To this end, we extend the range of $J$ to the full range observed in the simulations, i.e. $J \in [10^{-9},10^2]$.
Since the extended formulation \eqref{eq:mu_j_extended_rheology} is meant as an extension of the classical $\mu(J)$-rheology \eqref{eq:mu_j_rheology}, we keep the values of the previously determined coefficients $\mu_1, \mu_2$ and $J_f$ as reported in table \ref{tab:fitted_model_coefficients}.
This also effectively prevents possible overfitting.

The results are shown in figure \ref{fig:result_mu_viscousnumber_extended}, together with the existing parameterizations of the original model and the fits from \S\ref{sec:poly:fit}.
The obtained coefficients are given in the respective subcaption of the figures.
In all cases, the fit of the extended model (green line) is able to follow the shift to the creep regime and reproduces our simulation data very well, especially for the extended range $J\in [10^{-9},10^{-5}]$.
We also note that the curves of the extended model and the fit from \S\ref{sec:poly:fit} (orange line) collapse for $J > 10^{-4}$, where the extension term $\mu^c$ for the creep regime effectively evaluates to $\mu_1$ and thus reduces to the original model.
Analyzing the trend of the values determined for the two new parameters $\mu_0$ and $J_c$, we again notice a decrease in the friction coefficient $\mu_0$ with increasing polydispersity.
This decrease, however, is less significant than before for $\mu_1$ and $\mu_2$ and a difference of only around 10\% can be seen between the monodisperse case and the one with strongest polydispersity. 
Generally, $\mu_0$ is about three times smaller than $\mu_1$.
Determining the parameter $J_c$ faces similar challenges as discussed for $J_f$ before which thus shows no clear trend with polydispersity. It is obvious, however, that its value averages out around $10^{-6}$, which is more than three orders of magnitude smaller than $J_f$ and confirms the physical meaning of $J_c$ discussed above to describe the average value of $J$ for the {\em creep} regime.

Due to the observed marginal sensitivity of $\mu_0$ and $J_c$ on the polydispersity, and the general difficulty of measurements for the creep regime, we do not attempt to express a functional dependence on $\phi_m$ as in the previous section.
In order to obtain a general parameterization of the creep-extended model, we instead propose to use the following expressions, evaluated as the average of the fitted coefficients:
\begin{align}
\mu_0 &= 0.082 \pm 0.004 \label{eq:model_mu0},\\
J_c    &= 1.30\times 10^{-6} \pm 2.91\times 10^{-7}. \label{eq:model_Jc}
\end{align}

\begin{table}
	\centering
	\begin{tabular}{lcccccccccc}
	\toprule
		& \multicolumn{6}{c}{$R^2\big(\ln(\mu(J))\big)$} & \multicolumn{4}{c}{$R^2\big(\phi(J)/\phi_m\big)$}  \\ \cmidrule(lr){2-7}\cmidrule(lr){8-11}
		& \multicolumn{2}{c}{others} & \multicolumn{2}{c}{from \S\ref{sec:polydisperse_extension}} & \multicolumn{2}{c}{from \S\ref{sec:creep_extension}} & \multicolumn{2}{c}{others}  & \multicolumn{2}{c}{from \S\ref{sec:polydisperse_extension}} \\\cmidrule(lr){2-3}\cmidrule(lr){4-5}\cmidrule(lr){6-7}\cmidrule(lr){8-9}\cmidrule(lr){10-11}
		case & Boyer & Houssais & fit & correlation & fit & correlation & Boyer & Morris & fit & correlation \\\midrule
		\mono  &  0.625 & 0.726 & 0.761 & 0.759 & 0.994 & \textbf{0.992} & 0.995 & 0.989 & 0.996 & \textbf{0.996} \\
		
		\polya &  0.658 & 0.750 & 0.793 & 0.806 & 0.995 & \textbf{0.993} & 0.987 & 0.974 & 0.997 & \textbf{0.996} \\
		
		\polyb &  0.807 & 0.860 & 0.924 & 0.913 & 0.996 & \textbf{0.995} & 0.954 & 0.934 & 0.996 & \textbf{0.996} \\
		
		\polyc &  0.804 & 0.857 & 0.933 & 0.936 & 0.998 & \textbf{0.996} & 0.935 & 0.911 & 0.997 & \textbf{0.997} \\
		
		\bottomrule
	\end{tabular}
	\caption{$R^2$ values of different parameterizations of the rheology model, \eqref{eq:mu_j_rheology} and \eqref{eq:phi_j_rheology}, evaluated with respect to the simulated data for $J \in [10^{-9},10^2]$. 
	The polydispersity extension developed in \S\ref{sec:polydisperse_extension} features fits with coefficients from table \ref{tab:fitted_model_coefficients}, and the correlations \eqref{eq:model_mu1}-\eqref{eq:model_Kn}. 
	The creep extension, \eqref{eq:mu_j_extended_rheology}, in the current section re-uses these coefficients or correlations, respectively, and adds the coefficients from figure \ref{fig:result_mu_viscousnumber_extended} for the individual fits or the correlations from \eqref{eq:model_mu0}-\eqref{eq:model_Jc}.
	}
	\label{tab:extended_model_coefficients_r2}
\end{table}

We evaluate the performance of our creep-extended rheology model by computing the $R^2$ for the different empirical correlations over the entire range of $J\in[10^{-9},10^3]$. 
For that, we compare (i) \eqref{eq:mu_j_rheology} with the parameters of \cite{boyer2011}, (ii) \eqref{eq:mu_j_rheology} with the parameters of \cite{houssais2016}, (iii) \eqref{eq:mu_j_extended_rheology} with the parameters given in table \ref{tab:fitted_model_coefficients} and figure \ref{fig:result_mu_viscousnumber_extended}, and (iv) \eqref{eq:mu_j_extended_rheology} with the parameters given by correlations \eqref{eq:model_mu1}-\eqref{eq:model_Kn} and \eqref{eq:model_mu0}-\eqref{eq:model_Jc}.
The resulting $R^2$ values are given in table \ref{tab:extended_model_coefficients_r2}. 
In comparison to the existing model parameterizations of \cite{boyer2011} and \cite{houssais2016}, but also to the previously developed polydisperse model from \S\ref{sec:poly:model}, the creep-extended rheology outperforms all other available correlations. 
The fact that we observe an almost perfect match for both, the fit and the correlations, confirms the validity of our approach to account for polydispersity.

For completeness, we also show the $R^2$ values for $\phi(J)$ over the extended range of $J$, in contrast to the limited range used in table \ref{tab:fitted_model_coefficients_r2}.
From there, we see that the creep regime does not influence the predictive performance of the polydispersity-extended $\phi(J)$ model from \S\ref{sec:polydisperse_extension}, since it is the region of constant particle volume fraction and thus already covered by the model \eqref{eq:phi_j_rheology}. Overall, the parameterization of the creep-extended rheological model via the proposed correlations yields $R^2$ values larger than 0.992 for all here considered cases for both, $\mu$ and $\phi$, and without any further calibration. 
This is a significant improvement to the previous rheology model and its parameterizations.

\section{Conclusion}

In this work, we studied the rheological properties of polydisperse, densely packed sediment beds in a laminar shear flow through particle-resolved direct numerical simulations.
This was achieved by large-scale 3D simulation domains using an efficiently coupled lattice Boltzmann - discrete element method to fully resolve all relevant scales in space and time.
In particular, particle collisions are modeled by linear spring-damper models in normal and tangential directions, with a Coulomb-like friction model.
Additionally, a lubrication model is applied for short-range hydrodynamic interactions.
Four different sediment beds were created in a precursor simulation ranging from monodisperse to strongly polydisperse with a maximum to minimum diameter ratio close to 10. 
As a key feature, the non-uniformity of the sediment yields increasing values for the maximum packing fraction.
The beds consisted of up to 26000 particles, and the flow conditions were chosen to obtain several layers of mobile particles.
As such, the present simulations are one of the most extensive numerical studies on mobile polydisperse sediment beds.

From the simulation results, we obtained depth-resolved spatially and temporally averaged profiles of rheological quantities.
These enabled us to study the impact of polydispersity on the scaling of the macroscopic friction coefficient $\mu$ and the particle volume fraction $\phi$ as a function of the viscous number $J$, i.e., the $\mu(J)$-rheology. 
We compared our results to previous experimental studies of dense suspensions of neutrally buoyant spheres and sheared sediment beds of inertial particles and found excellent agreement for the monodisperse case. 
Owing to the wide value range of the viscous number, $J\in[10^{-9},10^3]$, and the highly-resolved data, we were able to enhance the $\mu(J)$-rheology and its parameterization for the effects of polydispersity and creeping flow. 
The effect of polydispersity has so far not been investigated for continuous grain-size distributions, and we addressed this issue by focusing on the frictional and hydrodynamic regimes. Based on our systematic simulation campaign, we derived an improved parameterization of the rheological model of \cite{boyer2011} that explicitly accounts for polydispersity. 
This was achieved by expressing the two coefficients $\mu_1$ and $\mu_2$, and the free parameter $K_n$ as functions of $\phi_m$.
The parameter $\phi_m$ is already present in the original rheological model and is here determined as the maximum observable packing fraction for a log-normal grain size distribution with a given variance, which determines the degree of polydispersity. 

The effect of creep has so far been reported in \cite{houssais2016} only, but this regime was excluded from the discussion of the rheology in this study. 
Our results confirm the existence of a creeping regime that is distinctively different from the well-known frictional and hydrodynamic regimes at higher viscous numbers \citep{boyer2011}. 
For vanishing shear, the macroscopic friction levels off to a quasi-static, creeping state that yields values of $\mu$, which are substantially lower than the frictional regime would suggest. 
This observation gave rise to the idea to enhance the $\mu(J)$-rheology to explicitly account for the creep regime following the argument of \cite{jop2005}. 
This was done at the cost of introducing two additional parameters.
However, we remark that these new parameters are physically based quantities related to particle properties as they express the quasi-static friction for the creeping state and the characteristic viscous number that describes the transition from the frictional to the creeping regime. 
These two parameters were determined by fitting the extended empirical correlation to our simulation data, and we found them to be less dependent on the maximum particle volume fraction. 
Compared to the frictional regime, the friction coefficient of the creeping regime is reduced by a factor of three.

Finally, our study demonstrates that particle properties that enter the $\mu(J)$-rheology framework may change the entire system's rheological properties. 
Since the scaling laws obtained so far involve several idealizations and particular choices for the sediment material used, more work will be needed to explore the effects of different particle and flow properties on the rheological behavior of sheared sediment beds.
This highlights another benefit of our simulation approach, where such changes can be made with ease, allowing for efficient parametric studies. \\

\noindent{\bf Supplementary data\bf{.}} \label{SupMat} Supplementary material and movies are available online. \\

\noindent{\bf Acknowledgements\bf{.}} The authors thank Morgane Houssais for sharing her data and gratefully acknowledge the Erlangen Regional Computing Center (\url{www.rrze.fau.de/}) as well as the Gauss Centre for Supercomputing e.V. (\url{www.gauss-centre.eu}) for funding this project by providing computing time on their supercomputers.  
\Rev{We thank three anonymous reviewers for their valuable comments that helped to improve the manuscript.}\\

\noindent{\bf Funding\bf{.}} B. V. gratefully acknowledges the support through the German Research Foundation (DFG) grant VO2413/2-1. U. R. gratefully acknowledges financial support by the Federal Ministry of Education and Research (BMBF) through the SKAMPY project, grant 01 ICH 15003 A, and by the Bavarian State Ministry of Science and the Arts through the Competence Network for Scientific High Performance Computing in Bavaria (KONWIHR). \\

\noindent{\bf Declaration of Interests\bf{.}}  The authors report no conflict of interest. \\

\noindent{\bf Author ORCID\bf{.}} C. Rettinger, \url{https://orcid.org/0000-0002-0605-3731}; S. Eibl, \url{https://orcid.org/0000-0002-1069-2720}; U. Rüde, \url{https://orcid.org/0000-0001-8796-8599}; B. Vowinckel, \url{https://orcid.org/0000-0001-6853-7750} \\

\noindent{\bf Author contributions\bf{.}} 
B. V. conceived the original idea.
C. R. performed the simulations and implemented the data analysis. 
C. R. and B. V. contributed equally in analyzing the data, developing the model extension, reaching conclusion, and in writing the paper.
S. E. implemented functionalities essential for polydisperse setups, assisted in plannings of the post-processing steps, and helped shape the research.
U. R. supervised the project.
All authors reviewed the final manuscript.
\\

\appendix
\section{Vertical size segregation}\label{sec:app1}

\begin{figure}
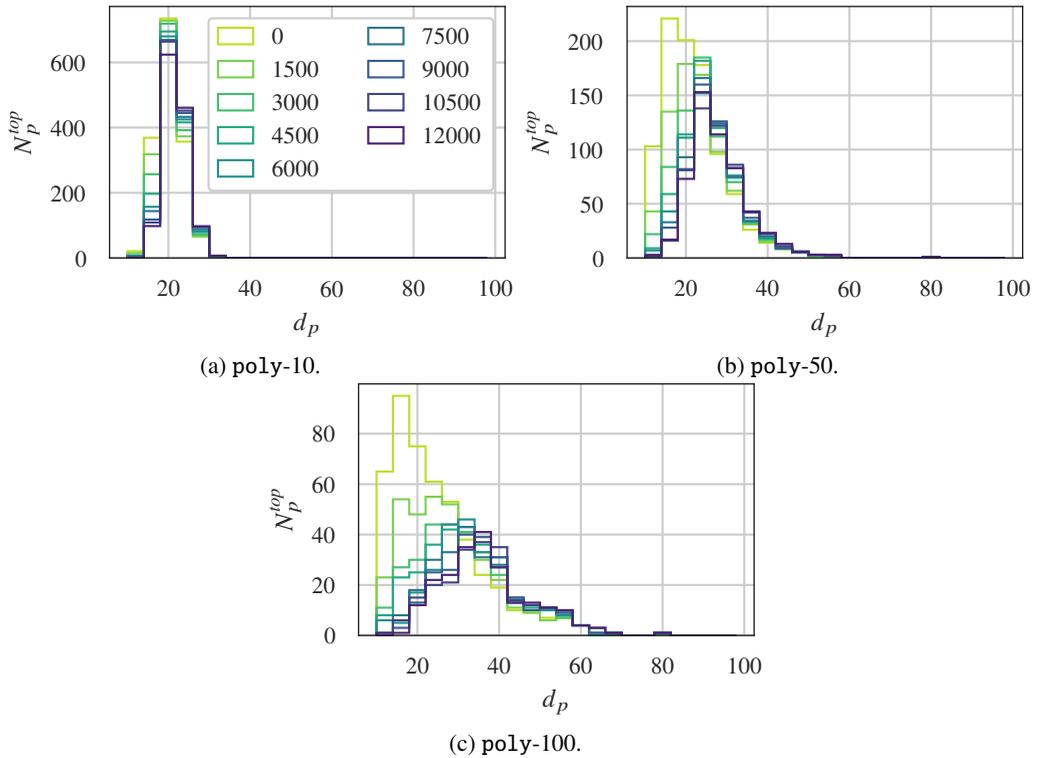

	\centering
	\begin{subfigure}{0.5\textwidth}
		\centering
		\input{figures/diameter_distribution_top_layer_var10.pgf}
		\caption{\polya.}
	\end{subfigure}~
	\begin{subfigure}{0.5\textwidth}
		\centering
		\input{figures/diameter_distribution_top_layer_var50.pgf}
		\caption{\polyb.}
	\end{subfigure}
	\begin{subfigure}{\textwidth}
		\centering
		\input{figures/diameter_distribution_top_layer_var100.pgf}
		\caption{\polyc.}
	\end{subfigure}
	\caption{\Rev{Temporal evolution of the top region's bed composition for the polydisperse cases, evaluated as the diameter distribution of the therein contained particles at distinct time steps $t/t_\mathit{ref}$.}}
	\label{fig:app:diameter_distributions}
\end{figure}

\Rev{
For polydisperse sediment beds that are exposed to shear stress, it is known that a vertical size segregation sets in \citep[e.g.][]{ferdowsi2017}. 
Consequently, larger particles move to the top of the bed while smaller particles descend to lower sediment layers.
A similar phenomenon, the \emph{brazil nut effect}, can be observed in dry granular beds subjected to vibrations~\citep{rosato1987}.

We study the dynamics of this vertical sorting by assessing the composition of the topmost layers of the bed.
To this end, we define that particles with a vertical center of mass position above $h^\mathit{top} = 15.5 \bar{d}_p$ belong to the bed's top region, which is roughly $2\bar{d}_p$ below the average sediment bed height $\langle h_b \rangle_t$, cf. table~\ref{tab:setups_evaluation}.
We then sort these $N_p^\mathit{top}$ topmost particles according to their diameters into bins of size $\bar{d}_p / 5$.
Evaluating the size distribution over time, we are able to investigate the size-based segregation in this top layer.
This evaluation is shown in figure~\ref{fig:app:diameter_distributions} for equally spaced time steps throughout the complete simulation, i.e., $t \in [0,12000]\, t_\mathit{ref}$.
Since such an effect is not present in the monodisperse case, we exclude it from these discussions.
}

\Rev{
In all cases, we see a qualitatively similar behavior.
The smaller size fractions, relative to the overall diameter distribution, decreases in number over time.
These particles, thus, move to lower layers of the bed and the smallest particles almost vanish completely from the top layers.
This process is initially very pronounced but then slows down gradually.
At the same time, the number of larger particles increases in the upper layer, although the absolute change is significantly weaker than for the smaller ones.
All these changes in the composition primarily happen during the initial stage of the simulation, so that a steady state develops after $t > 6000\,t_\mathit{ref}$.
This indicates that the fast segregation process, as described by \citet{ferdowsi2017}, is already completed. 
Therefore, we do not expect further strong morphological changes during the second half of the simulation from which we obtain the data for our evaluations, cf. table~\ref{tab:setups_evaluation}.
Since the present study focuses on sheared polydisperse sediments under well-developed conditions, this initial run-up phase was excluded from the statistical analysis presented in \S \ref{sec:rheology_monodisperse} - \S \ref{sec:creep_extension}.}

\bibliography{Library.bib}{}
\bibliographystyle{jfm}

\end{document}